\documentclass[a4paper,12pt]{article}

\usepackage[dvips]{graphicx,color}

\begin{document}
\title{EXPERIMENTAL EVIDENCE OF GIANT ELECTRON -- GAMMA BURSTS\\ GENERATED
BY EXTENSIVE
ATMOSPHERIC SHOWERS IN THUNDERCLOUDS}

\author{
A.V.~Gurevich$^1$, A.N.~Karashtin$^2$,
A.P.~Chubenko$^1$, L.M.~Duncan$^3$,\\
V.A.~Ryabov$^1$, A.S.~Shepetov$^1$,
V.P.~Antonova$^4$, S.V.~Kryukov$^4$,\\
V.V.~Piscal$^5$, M.O.~Ptitsyn$^1$,
L.I.~Vildanova$^5$, Yu.V.~Shlyugaev$^2$,
K.P.~Zybin$^1$}

\date{}
\maketitle

\begin{center}
{\it
(1) P.N. Lebedev Physical Institute, Moscow,  Russia\\
(2) Radiophysical Research Institute, Nizhny Novgorod, Russia\\
(3) Thayer School of Engineering, Dartmouth College, NH, USA\\
(4) Ionosphere Institute Almaty, Kazakhstan Republic\\
(5) Tien Shan Mountain Scientific Station of LPI, Almaty,
Kazakhstan Republic\\}
\end{center}

\begin{abstract}
The existence of a new phenomena -- {\it giant electron-gamma 
bursts} is established. The bursts are generated in 
thunderclouds as a result of the combined action of runaway 
breakdown and extensive atmosphere showers (RB-EAS). The 
experiments were fulfilled at the Tien Shan Mountain Scientific 
Station using EAS-Radio installation. This specially constructed installation consists 
of a wide spread EAS trigger array and a high time 
resolution radiointerferometer.
\end{abstract}

\section{Introduction}

A new physical concept of an avalanche type increase of a number
of relativistic electrons in gas under the action of the electric field 
was proposed by Gurevich, Milikh and Roussel - Dupre \cite{GMR}. The avalanche
can grow in electric field $E\geq E_c$. The field $E_c$ is almost an order of
magnitude less than the threshold electric field of conventional breakdown
$E_{th}$.  The growth of number of electrons with energies $\varepsilon >
\varepsilon_c \approx 0.1 - 1$MeV is determined by the fact that under the
action of electric field $E>E_c$ fast electrons could become runaways , what
means that they are accelerated by electric field $E$ as suggested by Wilson
\cite{Wil}. Due to collisions with gas molecules they can generate not only
large number of slow thermal electrons, but the new fast electrons
having energies $\varepsilon > \varepsilon_c$ as well. Directly
this process -- acceleration and collisions lead to the avalanche type growth
of the number of runaway and thermal electrons, which was called in \cite{GMR}
"runaway breakdown" (RB).
The detailed kinetic theory of RB was developed in \cite{RD}--\cite{GMZ03}.

In atmosphere, the critical field $E_c$ is $100\div150$ kV/m
and exactly these values of electric field are often observed during
thunderstorms \cite{Marsh}, \cite{MacG} . When the electric field in
thunderstorm cloud
reaches the critical value $E\geq E_c$ every cosmic ray secondary electron
(its energy $\varepsilon>1$MeV) initiates a micro runaway breakdown.
It serves as a source of intensive ionization of air and manifests
itself in a strong amplification of $X$ and $\gamma$ -rays emission in
thundercloud \cite{GMR},\cite{AVG},\cite{DWY}.

This emission was observed at airplanes \cite{McC}, baloons
\cite{Eack}, and in the mountain experiments \cite{Ch} -- \cite{18a}.

Extensive atmospheric showers (EAS) are generated by a high energy
$\varepsilon$ cosmic ray particles. EAS are accompanied by a very
strong local growth of cosmic ray secondaries number and
$\gamma-$emission in a high energy range $\varepsilon \sim 10
\div 100 MeV$. This flux interacts effectively 
with thundercloud electric field  if runaway breakdown conditions
are fulfilled (RB-EAS interaction). The
energy obtained by fast electrons from thundercloud electric
field due to the RB-EAS combined effect can serve for
the explosive generation of a very large number of a newborn electrons
and gamma quanta (see
accompanying theoretical paper \cite{GDMZ2}).

The experimental study at the Tien Shan Mountain Scientific Station
of this phenomena is our  goal. 

At the  station a special system of gamma-spectrometers 
exist for "thunderstorm" measurements. It
is based on the  Geiger-Muller counters SIG5 and
{\it NaJ} detectors. The system permit to measure intensity of
electrons, X and gamma quanta in wide energy range. A
detailed description of "thunderstorm" detector complex can be
found in \cite{Ch},\cite{Ch2}. 

For the present work a
special air shower trigger array was created. The array system
has to fix the EAS coming on an extensive region with high
accuracy. The system consists of a set of shower electrons and
$\gamma-$quanta detectors (based on proportional Geiger-Muller
counters) widely spread over Tien Shan Station territory.

The RB-EAS combined action lead to a strong radio pulses emitted
from thundercloud \cite{GDMZ}. A specially constructed radio
installation was used to study the short electromagnetic
pulses of this radio emission \cite{G03}. The installation allows to
determine pulse form, its maximal intensity and arrival direction
(inclination and azimuth angles). The moments  of EAS coming
is fixed by piping the trigger signal from shower array to the
radio installation. That allows to study the EAS and radio
emission simultaneously.

The results of measurements  are
presented. More then 150 simultaneous EAS-radio events are
observed. The analysis of the obtained experimental data allows
to establish the existence of a new phenomena -- {\it giant
electron -- gamma bursts} generated in thunderclouds.

\section{The cosmic showers trigger array}

The cosmic showers trigger array has to fix coming EAS and inform
about it radio installation piping there trigger signal. The array is
fixed around Tien - Shan Mountain Scientific Station situated
3340 m high above see level ($43^o02'$N,  $76^o56'$E)

\noindent {\bf The basic design principles.}

The basis of the trigger array topology is the principle of three-fold
signal coincidences from a set of detectors placed in the vertexes of a
triangle. The whole system was build as an aggregate of such triangles.
The common trigger signal is the logical \emph{OR} sum of signals, generated
independently by elementary triangle parts.

For the EAS particles detection in trigger array are used the SI5G type
self-quenching ionization counters which permit to have a high amplitude
signal at relativistic charged particles and high energy gamma quanta
passage. The counters are tolerant to sharp changes of weather conditions.
The registration efficiency of SI5G counters for relativistic
electrons  is high ($>$ 80\%). For gamma quanta it is much lower and depend
on quanta energy. The detailed calibration measurements of SI5G counter
efficiency for registration of gamma quanta in our conditions were carried out
in \cite{Ch2}. The characteristic efficiency is of the order of 1\%. Note,
that it could be noticeably increased due to photoelectrons emitted
from the aluminum walls of the counter box.

The counters are placed inside the boxes made of a 2~mm thick aluminum,
which assures a proper shielding from electromagnetic interference. Each
detector box contains 18--20 counters. 
The
sensitive area of each SI5G counter being 0.033~m$^2$, the total area of a
detector box is around 0.6~$m^2$.
The detector is placed in a special little housing
construction \cite{Ch}. 

Pulse signals from the all distant detector points are transmitted to the data
collection centers through the screened electrical cables. To avoid the damage
of semiconductor electronics in the moment of lightning discharge, the cables
are conjugated both with the detector and with registering apparatus throughout
the vacuum tube based circuits (see \cite{Ch},\cite{Ch2}).

\noindent {\bf Layout of the shower trigger array.}

\begin{figure}[p]
\begin{center}
\includegraphics[height=100mm]{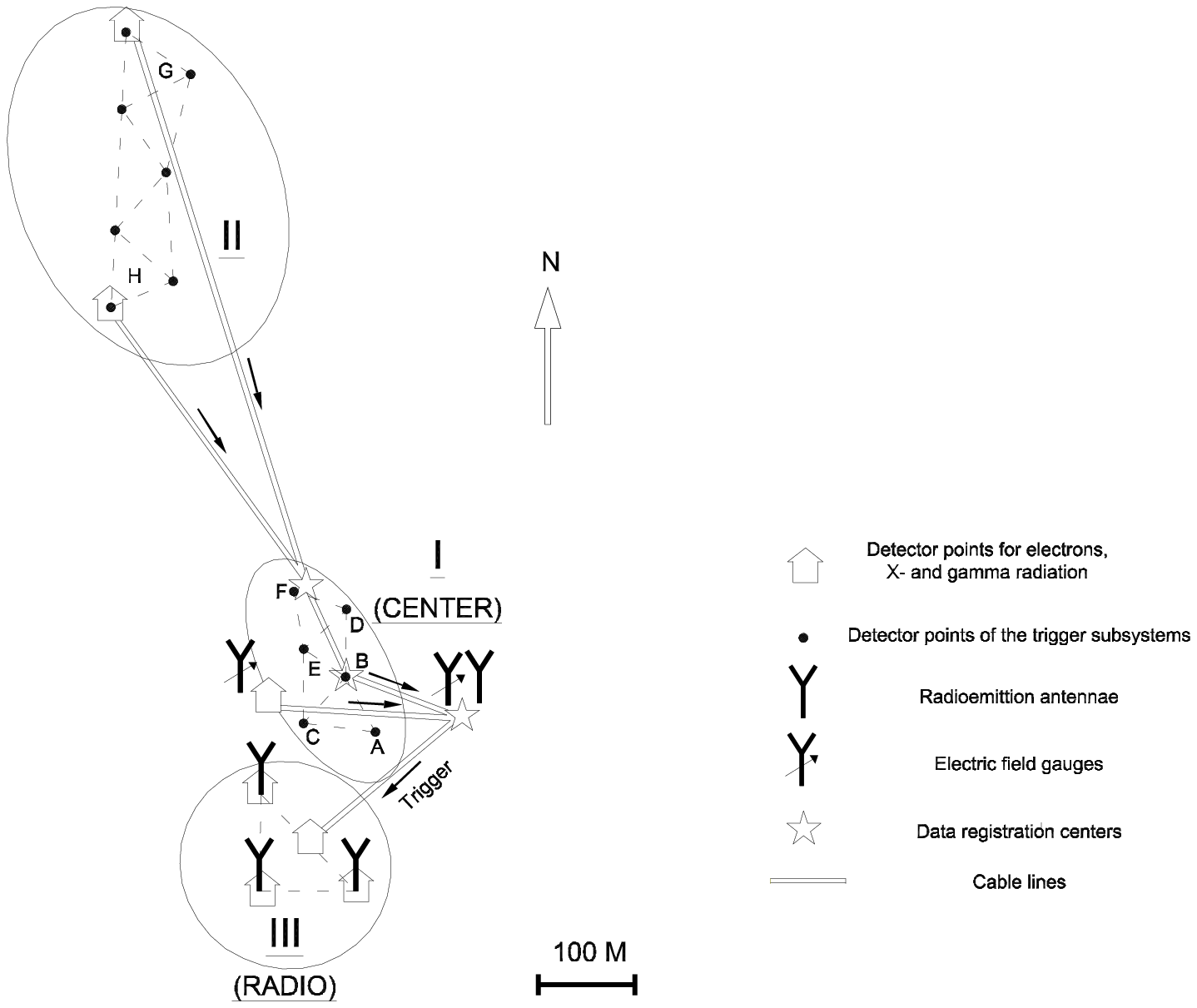}\\
\vspace{10mm}
\includegraphics[height=80mm]{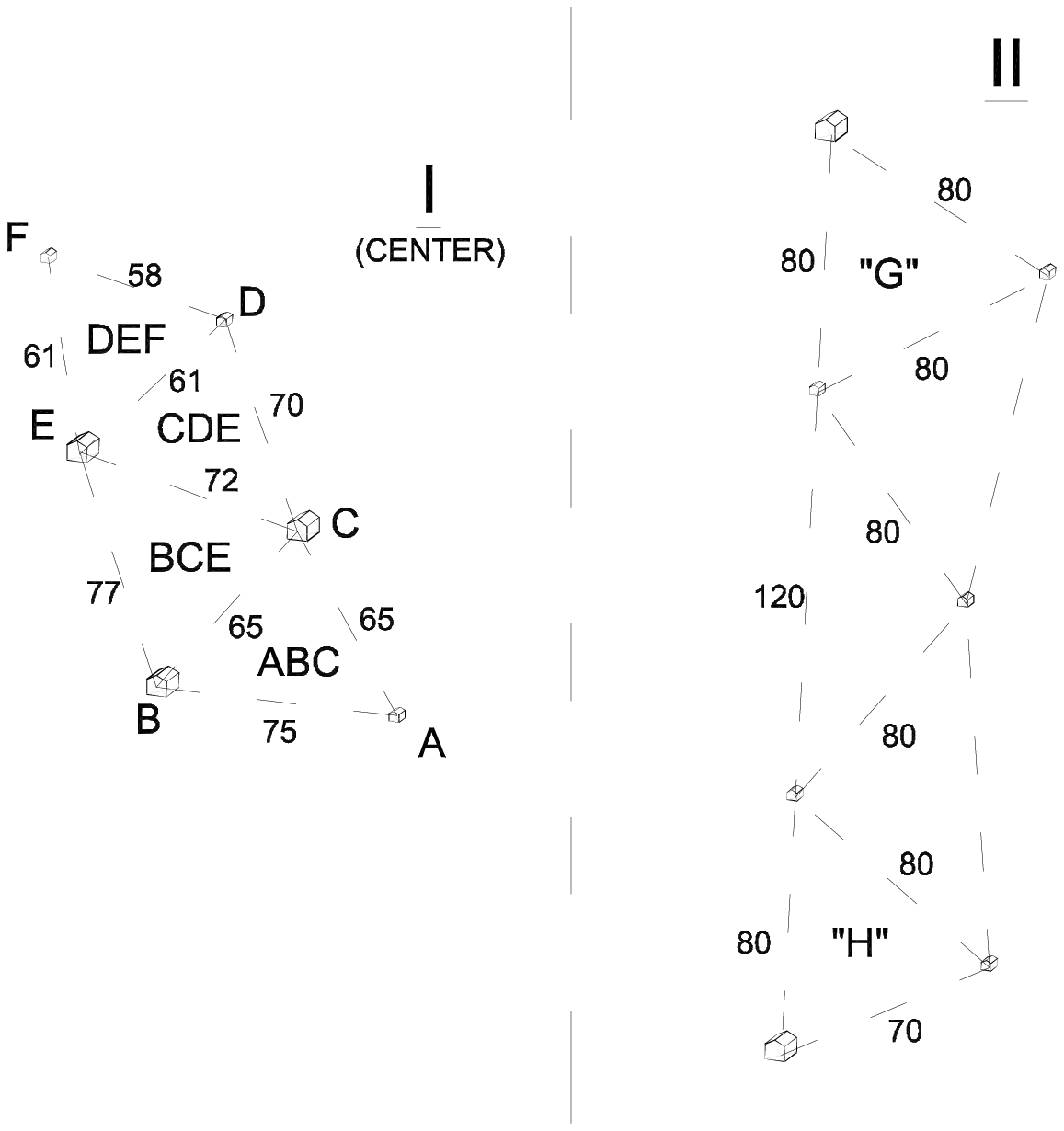}
\end{center}
\caption{Up -- Common layout of the shower trigger array at
Tien Shan station (3340~m~a.s.l.) in Summer~2003. Detector
subsystems are encircled with solid ellipses. Arrows shown near
the cable lines indicate the direction of information transfer.
The subsystem \emph{II} is raised up to 200~m above the common
Tien Shan station's level.}

Down -- layout of the separate trigger
subsystems \emph{I-II}. Numerals shown near the lines represent
the length of corresponding  triangle edges (measured in meters),
the letters mark separate detector points and corresponding
coincidence types.
\label{Lay}
\end{figure}

Current status of the shower trigger array is shown in Fig.~\ref{Lay}. The
array consists of two independent subsystems marked with Roman numerals on
the figure; each subsystem, in its turn, --- of a set of elementary triangles.

As it is seen from Fig.~\ref{Lay}, the total area being covered with
trigger array's perimeter exceeds 0.1 square kilometer.

The size of the elementary detector triangles was determined by such a way to
achieve the rate of triple coincidences about one per 10 seconds.
Preliminary measurements have shown that
our first triangle ABC having the edge sizes
65~x~75~x~65~m$^3$ satisfy this requirement. This triangle was selected as a
basic for the trigger array constructing.
Fig.~\ref{Lay}(down) represents on a large scale the mutual location of
detector points in the  trigger subsystems. The
numbers shown near the edges of elementary triangles mean their length in
meters.

\noindent {\bf Trigger data requisition.}

The shower trigger array (so as the whole "thunderstorm" detector system) has
several data registration centers connected with each other and with the
trigger detector subsystems by the means of cable lines. Topology of the
data transfer cables is shown in Fig.~\ref{Lay}.

One of the registration centers --- the main center --- has a cable connection
also with the system of radio emission registration; it is the same place where
the common shower trigger signal is elaborated from coincidence signals of
separate elementary triangles. The common trigger signal is transmitted to the
radio registration system through the special cable; in the moment of its
appearance information about the coincidence type (namely, identifiers of the
all detector points whose pulses were coinciding with the trigger) is written
by a registration computer. Later these data are used to estimate roughly the
size of a registered shower.

All the detector points of subsystem \emph{I} (it is also denoted as
\emph{CENTER} in Fig.~\ref{Lay}) have a direct connection with the main
registration point and their pulses are transmitted to that point immediately.
This circumstance permits to analyze the trigger coincidence type in
particularly and to define precisely the intensity of \emph{ABC}, \emph{BCE},
\emph{CDE} and \emph{DEF} triggers coming from the \emph{CENTER} trigger
subsystem.

The signals of the subsystem \emph{II} are analyzed in the peripheral
registration point and just the ready coincidence pulses from its basic
triangles \emph{G} and \emph{H} are transmitted to the main center (see
Fig.~\ref{Lay}).

In the main registration center any combination of the \emph{ABC}, \emph{BCE},
\emph{CDE}, \emph{DEF},  \emph{G},  \emph{H} signals is used to generate a
common trigger pulse for radio emission registration system. 
The trigger signal is formed as a coincidence of
signals within $5 \mu$s duration from every three modulus connected in a
triangle (ABC for example). That is why the trigger time accuracy of the array
is $5 \mu$s. Its  stability was tested with a precision scintillation counter.

\noindent {\bf Calibration of the array during quiet time}

During quiet time (not thunderstorm days) trigger array was carefully
calibrated. It was established that in average  the full array 
was 
registering a shower every 2.45 second (Fig.~\ref{evnum}). Calibration was proved
by measurements at triangle ABC done simultaneously by SIG5G and 
scintillation counter
counters. A detailed numerical simulations was fulfilled, which demonstrated
a reasonable agreement with observations.

It will be shown
below that the work of a trigger system in thunderstorm period (which is
of our interest) and in quiet days
{\it is significantly different}.  That is why the detailed
description of a quiet time array calibration is omitted here. It will be
presented in a separate paper.

\section{Radio registration system}

\noindent {\bf Receiving antennas}

\begin{figure}[p]
\begin{center}
\includegraphics[width=21cm]{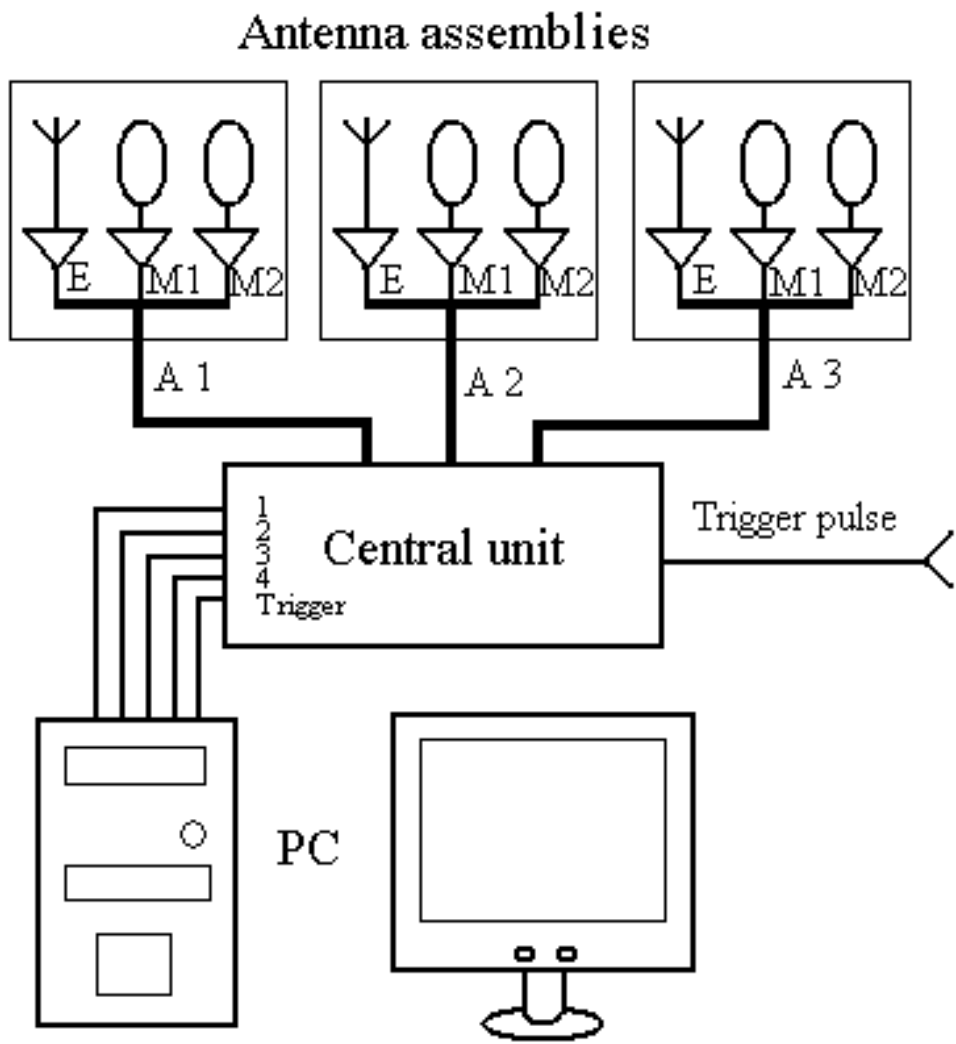}
\end{center}
\caption{Block diagram of the HF installation}
\label{fig:instr}
\end{figure}

Radio measurements were carried out using specially designed
installation (analogous to described in \cite{G03})
for short electromagnetic pulse observations in the
frequency range from 0.1 to 30 MHz. This installation contains
three spaced receiving antennas connected to the receiving
apparatus (central unit). It allows signal arrival angle
determination using correlation technique as well as waveform
recording with high temporal resolution (16 ns). Its block diagram is
shown in the Fig.~\ref{fig:instr}.

Each receiving antenna of the installation (antenna assembly)
actually consists of three individual antennas---two mutually
perpendicular loop antennas for horizontal magnetic field
measurements, and an End-Fed antenna to measure vertical electric
field. All three antennas are active and include transistor
amplifiers. This allowed to reduce the size of the assembly
essentially and to reach uniform gain-frequency characteristics
for all antennas in the wide frequency range.

Antenna for magnetic field measurements is a screened vertically
sited rectangular loop. Loop and prime amplifier screening is
needed to provide antenna sensitivity only to magnetic field, in
other case its directional pattern becomes non-symmetric and
frequency dependent. To equalize frequency-response of the loop it
should be loaded by small resistance. Thus it is attached to the
amplifier input through a step-up transformer made on long lines
working as current transformer.

Electric antenna is an electrically short rod attached to the
matched transistor amplifier. Its frequency-response equalization
is achieved using principally capacitate character of the
electrically short pole impedance as well as of transistor input
impedance in the frequency range concerned. So, antenna--amplifier
connection is a frequency independent divider with a ratio of
antenna capacity to amplifier capacity as a transfer coefficient.
Antenna capacity is increased by a plate tip at the end of the rod
to increase the transfer coefficient. The increase of antenna
capacity extends the antenna frequency range to lower frequencies
also. Low boundary of the antenna frequency range is determined by
the response time that is equal to the product of the antenna
capacity and an input resistance of the amplifier. High input
resistance of the amplifier leads to the growth of low frequency
noise and assigns a risk of input transistor damage due to
breakdown in strong electric fields. Therefore the input
resistance was chosen in accordance with low boundary of the
antenna frequency range (100 kHz approximately) to be about 1~MOhm. 
Breakdown protection of the amplifier is provided in three
stages by high-voltage feed-through capacitor, gas-filled excess
voltage preventer, and two parallel-opposition semiconductor
diodes successively mounted at its input.

\noindent {\bf Data registration and storage}

All three antenna amplifiers outputs from each antenna assembly
are connected to the central unit by coaxial cables of equal
lengths. Separate cables are used to provide antenna amplifiers
power supply through appropriate bypasses. The central unit
provides switching and selection up to 4 receiving elements
(antennas) most suitable under specific conditions of the
experiment. It serves also for preliminary analog data reduction,
namely, reduction of the dynamic range of received signals to
analog-to-digital converter (ADC) input range to prevent loss of
information, and both high-pass and low-pass filtering to decrease
low-frequency noise and to prevent spectral aliasing during data
sampling.

After the central unit chosen signals are digitized by 4-channel
radio receiver based on the personal computer (PC) using two
synchronized two-channel ADC boards with sampling rate up to 60
MHz and 256 MB per board internal memory. High sampling rate leads
to high data traffic that exceeds the ability of PC bus. Therefore
pulse operation mode is needed with fast data storage into
internal buffer area following by slower data transfer to the PC
initiated by trigger pulse. ADC internal buffer memory allows to
record up to 1 second sessions with maximum sampling rate of 60
MHz.

Radio receiver is operated in the pretrigger mode allowing signal
recording before and after the trigger pulse. It can be triggered
by external TTL signal as well as internally. In the last case
triggering occurred when an input signal of one of the ADC
channels exceeds a preset threshold.

The whole installation is supplied by UPS (Uninterruptible Power
Supply) with enhanced up to several hours off-line operation
capability (using external accumulator of higher capacity).

Antenna units (assemblies) are mounted at the Tien Shan Scientific
Station at the altitude of about 3340 m above the sea level. They
form direct angle triangle with perpendicular bases of about 78 m
in the $350^{\circ}$ azimuth direction with elevation angle of
$-7^{\circ}30^{'}$ (below the horizon) and about 74 m in the
$80^{\circ}$ azimuth direction with elevation angle of
$-2^{\circ}25^{'}$ (above the horizon).

An additional receiver operated at about 250 MHz is used as a
separate facility to observe VHF radiation from the lightning.
This receiver has filter bandwidth of 6 MHz, its output is
detected with time constant of 100 $\mu$s or 10 $\mu$s providing
correspondent time resolution. Its data acquisition system can
also record EAS trigger pulses and static electric field using
additional channels.

\section{Results of measurements}

\noindent {\it Quiet time}

\begin{figure}[p]
\begin{center}
\includegraphics[width=16cm]{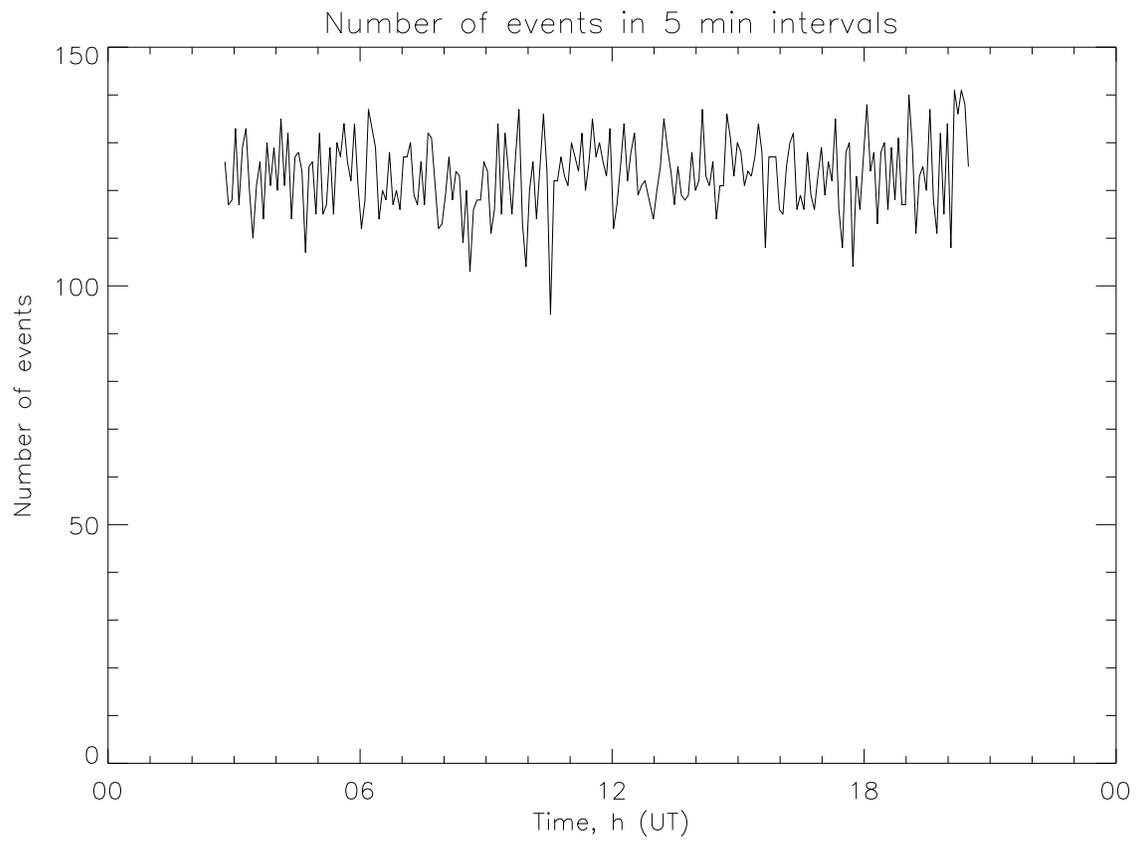}
\end{center}
\caption{Trigger array signals in quiet time}
\label{evnum}
\end{figure}

\begin{figure}[p]
\begin{center}
\includegraphics[width=15cm]{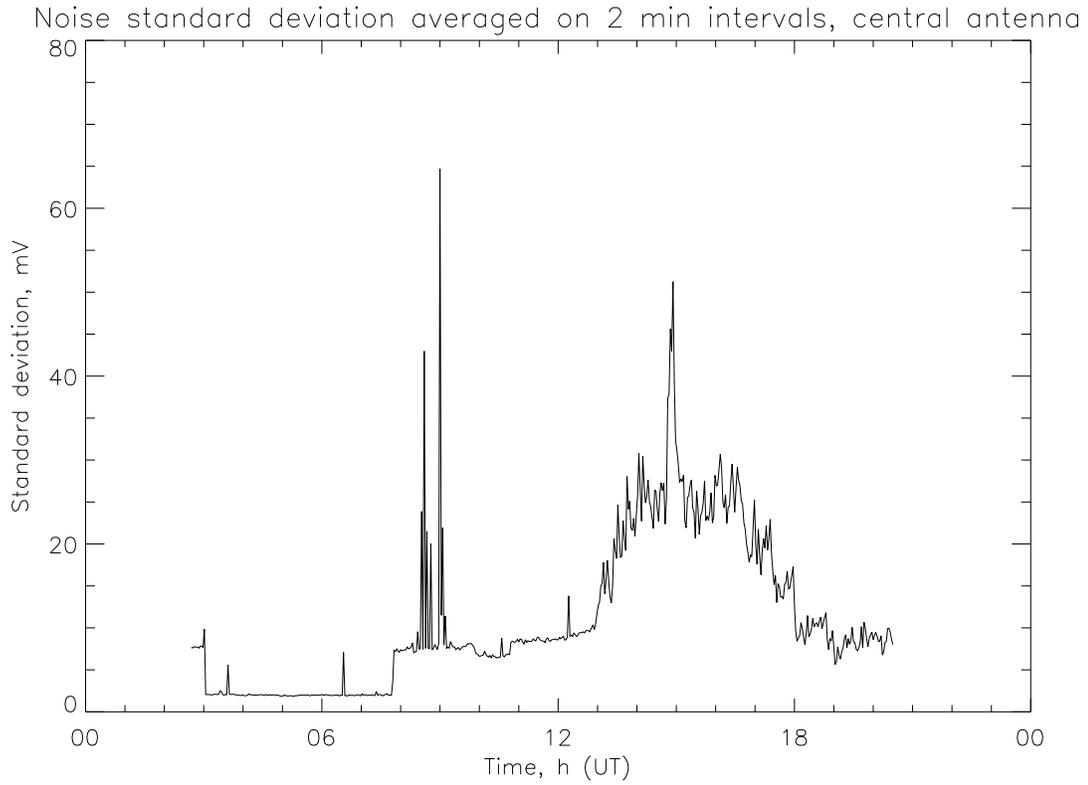}
\end{center}
\caption{The noise is minimal
$\sim 2 - 3$mv 03 -- 08 h (UT) or at 10 --15 h local summer time (LST).
During 08 -- 12 h (UT) or 15 -- 19 (LST)
the mean amplitude of the average noise electric field is
4 -- 5 times  larger (about 8 -- 10 mv) due to the work of Alma - Ata
radio stations. In the night time  the noise was
much higher (about 30 mv) due to disappearance of D-E layer absorption.
}
\label{noise}
\end{figure}

Observations were carried out from July 19 till October 09, 2003
almost continuously. The radio
receiving system was triggered by the EAS facility. Mean time
interval between triggers was about 10 s in the beginning of
observations decreasing later to  2.45 s due to upgrade of EAS
facility and enhancement of its effective area (Fig.~\ref{evnum}).
Recording time was
slightly above 100 $\mu$s with 83.2 $\mu$s pre-history. Data
sampling rate of 60 MHz was used. Short radio pulses with few
hundred nanosecond length correspondent to EAS's were searched.
Wide band radio interferometer in frequency range 0.1 -- 30 MHz was used
in our experiment. The electromagnetic noise in this frequency range
is always rather strong. The radio noise spectrum has well defined maximum 
of the order or less than 0.2 MHz. This frequency is
small compared with the frequency of registered signals ($\approx 3$ MHz). 
That's why only noise intensity  is significant. It is shown in Fig.~\ref{noise}. 
Strong changes of noise with time is seen

We note, that the mean electric field amplitude value (in mv/m) approximately
1.5 times larger than the potential (in mv) at antenna rode, shown at the
figure.

\begin{figure}[p]
\begin{center}
\includegraphics[width=15cm]{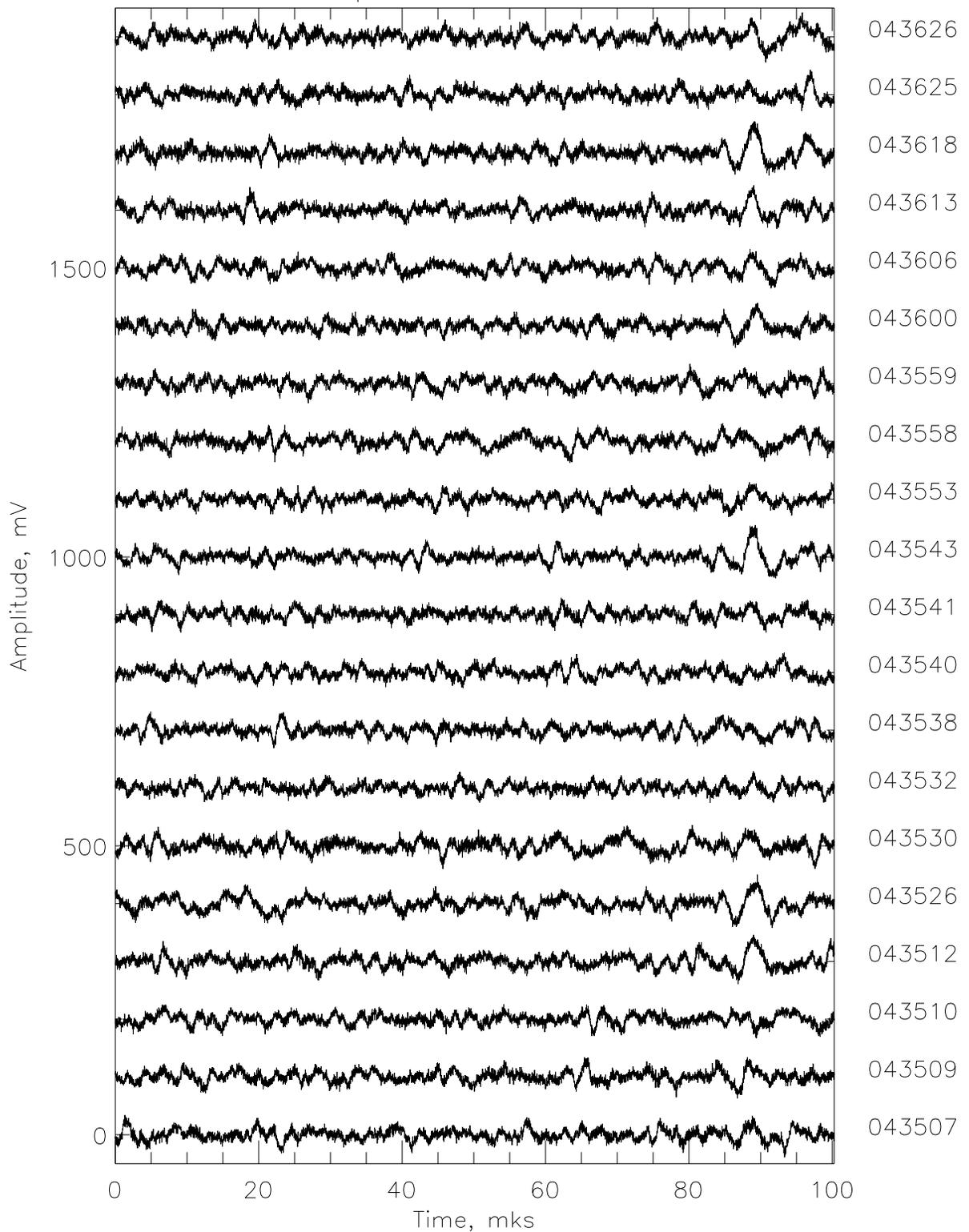}
\end{center}
\caption{An example of hundred microsecond quiet time radio record triggered by EAS.
Trigger center - 83.2 $\mu$s, trigger window - 5 $\mu$s. Twenty records of the central 
antenna potential are presented, mV. Every next record is 
shifted on 100 mV. The sampling time is presented in the right column. 
For example, in the number 043507 04 means the hour, 35 - the minute, 07 - the second.}
\label{exampq}
\end{figure}

The general result of the observations is that the {\it short radio
pulses under quiet conditions are absent}.
An example of this type regular record
in quiet time is shown at Fig.~\ref{exampq}. One can see, that the trigger electric
circuit signals lead to the noise amplification around trigger point
82.2$\mu s$.

\noindent {\it Far thunderstorm}

Few records were done for far thunderstorms
in the autonomous mode with triggering from
the output of one of the receivers as it was described in
\cite{G03}. The record length was 750 ms
with 500 ms pre-history. The same behavior of the radio signal
during lightning initiation was observed as it was reported in
\cite{G03} for lightning in the middle
Russia far from mountains. Every lightning began with a short pulse
and there were no other pulses during 0.5 sec before the first one.

\noindent {\it Close thunderstorm}

\begin{figure}[p]
\begin{center}
\includegraphics[width=15cm]{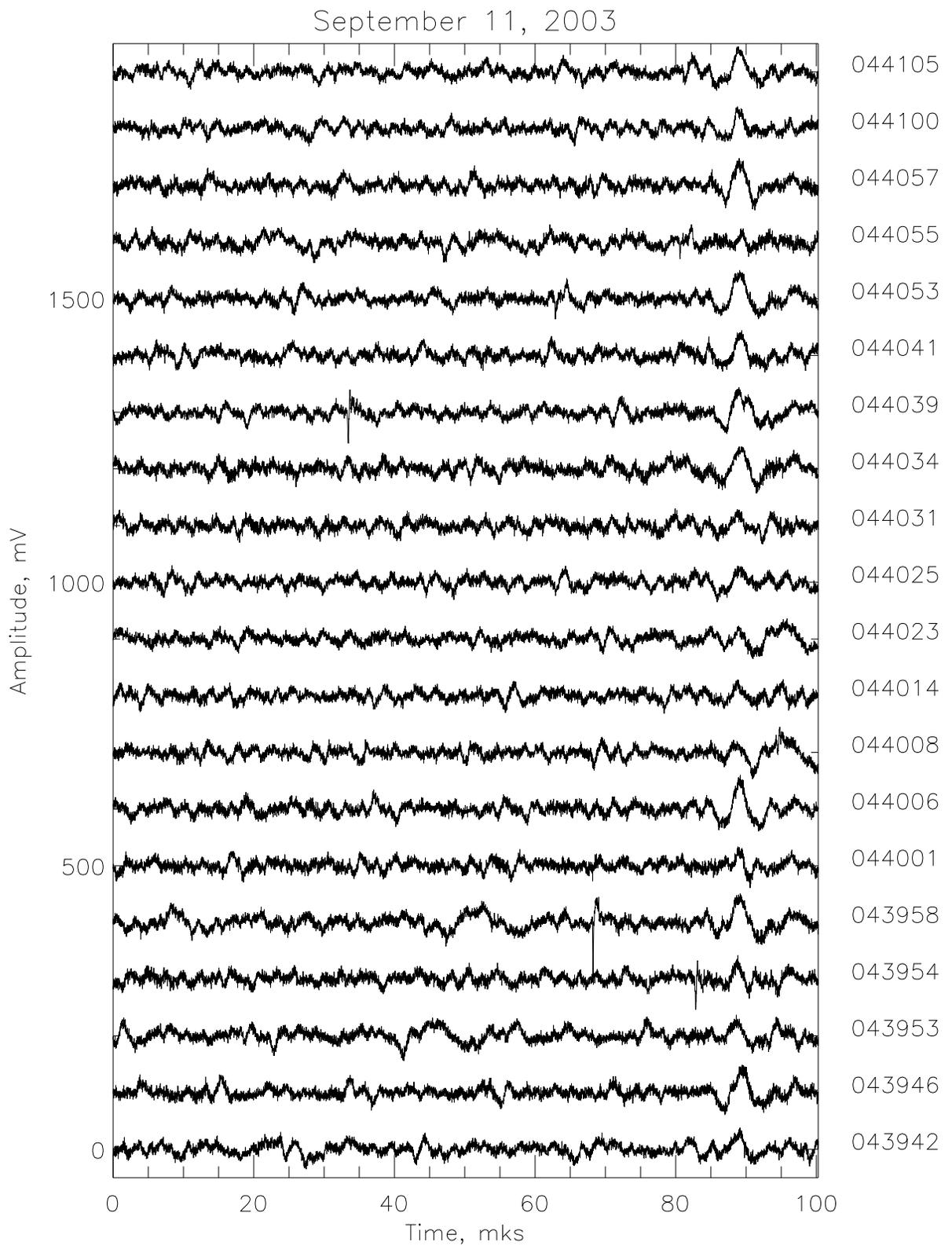}
\end{center}
\caption{An example of hundred microsecond thunderstorm time radio record triggered by EAS.
Trigger center - 83.2 $\mu$s, trigger window - 5 $\mu$s. Short 
radio pulses are seen in records 044039, 043958, 043954.
}
\label{exampt}
\end{figure}

Unfortunately, during the period of the observations there were no
strong  thunderstorms  directly above the station.
Two thunderstorms on July 23 and September 11, 2003 occurred in
its close vicinity with lightning at the line-of-site range.
An example of records in thunderstorm conditions is given in 
Fig.~\ref{exampt}. Short radio pulses are seen.

The noise amplification around trigger point in
thunderstorm time become  stronger

During  thunderstorms a number of short electromagnetic
pulses were recorded.
An example of the record with such a pulse is
presented in the Fig.~\ref{fig:record}. Four shown data records
correspond to North, Central, Central, and East electric field
antennas. Central antenna signal was recorded twice by different
channels of different ADC boards to diminish possible errors in
the determination of signal delays between antennas for more
accurate arrival angle calculations.

Some records show two short pulses in the 100 $\mu$s interval of
the record. Besides, on July 23 there were a few records of strong
electromagnetic signals of different structure possibly
correspondent to radiation from electric power line.

The  results of short electromagnetic pulse observations
for thunderstorms on July 23 and September
11, 2003. are summarized
in the Tables~1 and~2 respectively. The
time and number of event, delay of the trigger relative to radio
pulse, and radio pulse arrival angles are given in the Tables. The
delay is positive if radio pulses were observed before trigger and
negative while after it. Arrival angle is given as an inclination
angle from the vertical and an azimuth from the North
of the direction to the radio source.

The summary of the short pulse arrival angles is shown in the
Figures~\ref{fig:rjul} and~\ref{fig:rsep} as inclination--azimuth
plots for July 23 and September 11, 2003 thunderstorms
respectively. It is clearly seen that in both cases most radio
pulses came from the horizon direction. Note, that on September 11, 2003
there were several pulses which came from the direction near the vertical.

\begin{figure}
\begin{center}
\includegraphics[width=15cm]{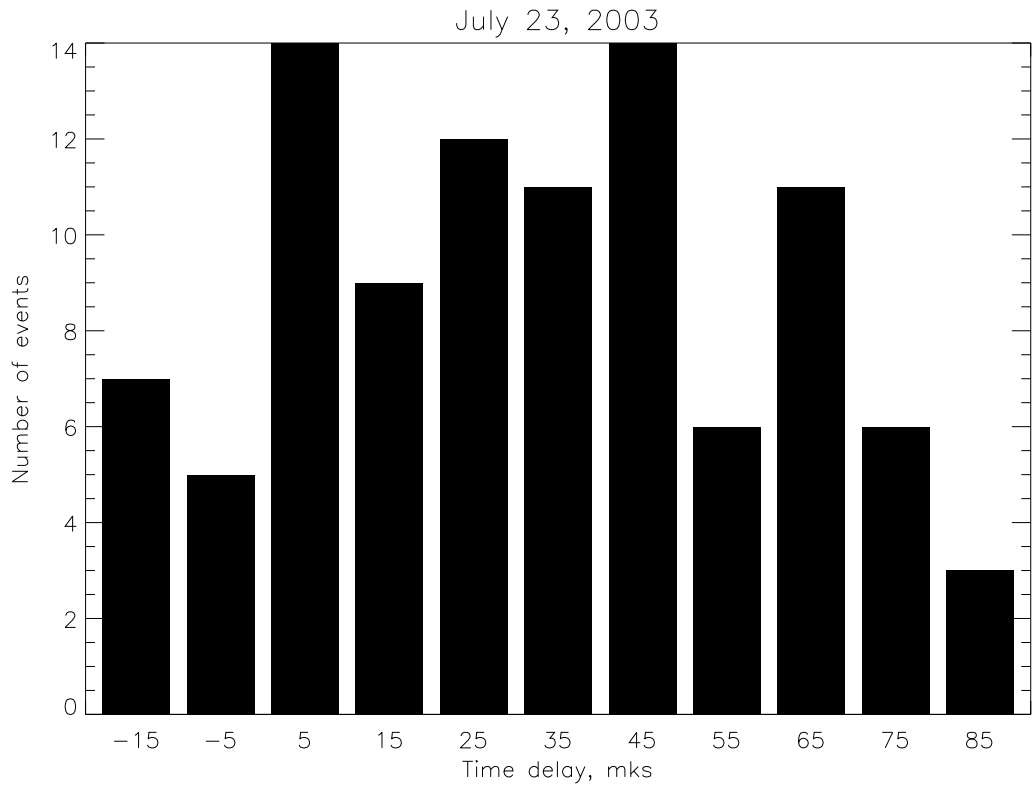}
\end{center}
\caption{Histograms of the delays between radio pulse arrival and EAS 
during July 23, 2003 thunderstorm. Time delay in $\mu$s.}
\label{fig:hjul}
\end{figure}

Histograms of the delays between radio pulse arrival and EAS
triggers are shown in the Figures~\ref{fig:hjul}
and~\ref{fig:hsep} for July 23 and September 11, 2003
thunderstorms respectively. Positive delays correspond to radio
pulses arrived before trigger pulse. It is seen that delays are
near equally distributed over the recorded interval.

The radio pulses generated during two observed thunderstorms
have similar main features. Below we will analyze in details mainly the
thunderstorm 11 September. The trigger array at this storm was
wider: the radio pulses I and II in Fig~\ref{Lay} for September 11
were used and only I for July 23.
The array calibration at September 11 was totally completed and
the time interval between triggers was stable (Fig.~\ref{evnum}).
Only three records show two short pulses at 100 $\mu$s interval
and a few were observed after trigger.

\section{Discussion}

\noindent {\bf I. Radio pulses}

{\it RB - EAS combined action} The process which is called RB - EAS
combined action
take place when high energy cosmic ray particle cross the thundercloud.
The thundercloud is supposed to be in RB state, what means that the maximal
electric field in thundercloud is close to  $E_c$ :  $E_m\sim 1.0 - 1.2
E_c$. Exactly these values  of maximal electric field are observed in
thundercloud at balloon experiments \cite{Marsh}. Large scale thundercloud
electric field is directed close to vertical $z$ \cite{MacG}.

The secondary electrons  of EAS are multiplied and accelerated in
thundercloud due to RB process \cite{RD} - \cite{UFN}. In the same time
a very large number of thermal electrons is generated  due to air
moleculas ionization  by high energy electrons \cite{GMZ03}. The thermal
electrons move in the air under the action of thunderstorm electric
field forming an electric current pulse. The current pulse direction
is determined by the direction of thundercloud electric field: it could
be positive or negative depending on field direction (up or down).

{\it The current pulse } is unipolar. Its growth time $\Delta t$ is determined
by RB process $\Delta t \approx l_a/c$, where $l_a$ is characteristic
scale of relativistic electron avalanche. At thundercloud heights
not far from the Station (4 -- 5) km
$l_a\approx 60 - 100$ m \cite{GMR},\cite{UFN}. The pulse decay is determined by thermal
electrons disappearance due to triple collisions attachment in air.
The characteristic  time of triple attachment
at heights 4 -- 6 km
$\Delta t_{at} \approx 100$ ns \cite{GBM},\cite{BAZ}.

\begin{figure}
\begin{center}
\includegraphics[width=15cm]{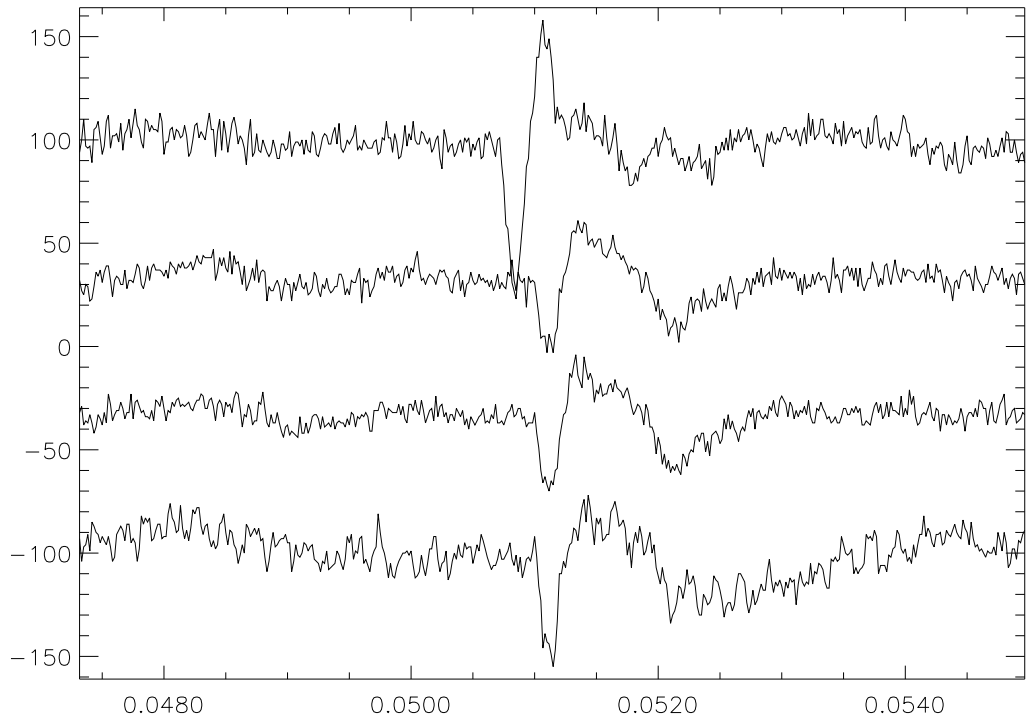}
\end{center}
\caption{An example of the short pulse record obtained during September 11, 2003 
        thunderstorm (event 1258).
         Time is in milliseconds from the beginning of the record.
         Amplitude is in mV. The short pulse is seen at about 0.051 ms.
         Trigger is at 0.0832 ms, so the delay is about 0.03 ms.
         }
\label{fig:record}
\end{figure}

{\it The form of radio pulse}, generated by unipolar current pulse is
bipolar \cite{GDMZ}. Its characteristic time scales, determined by current
pulse are $\Delta t_{at}\sim $100 ns and $\Delta t\sim l_a/c\approx
200 \div 300$ ns. The full width of a pulse $\sim 0.5 \mu$s
Exactly these time scales determine the frequency range and the
form of observed in our experiment radio pulses (Fig.~\ref{fig:record}).

\begin{figure}
\begin{center}
\includegraphics[width=15cm]{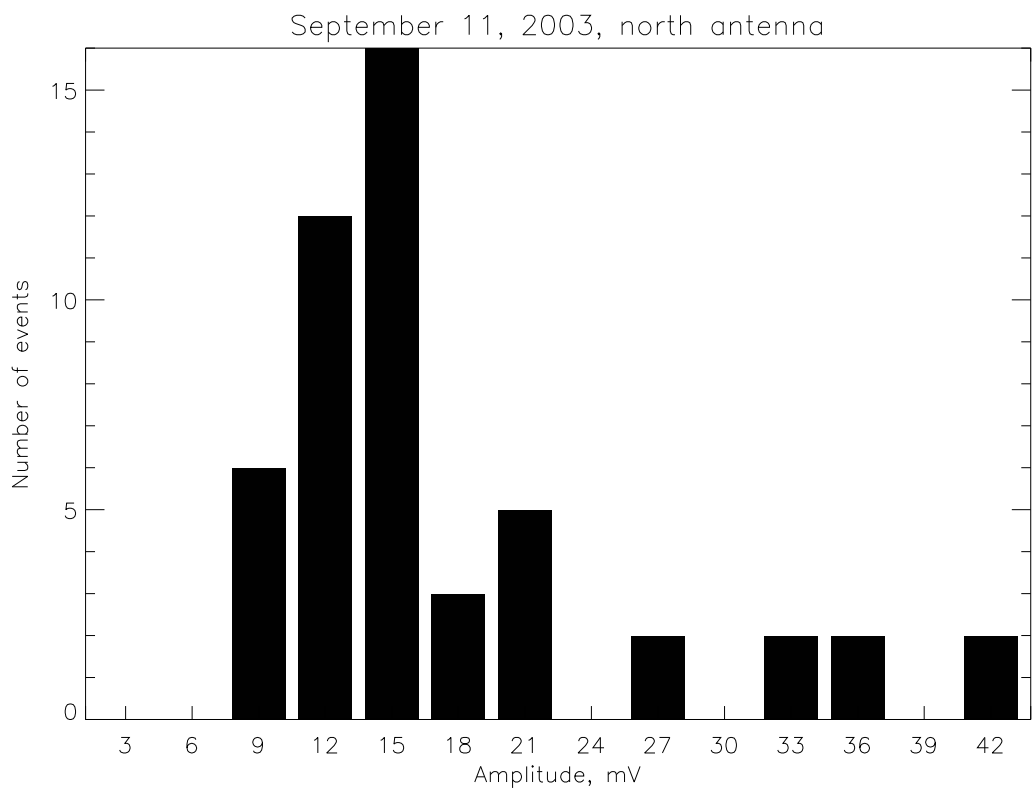}
\end{center}
\caption{Histogram of pulse amplitude.}
\label{hamp}
\end{figure}

\begin{figure}
\begin{center}
\includegraphics[width=15cm]{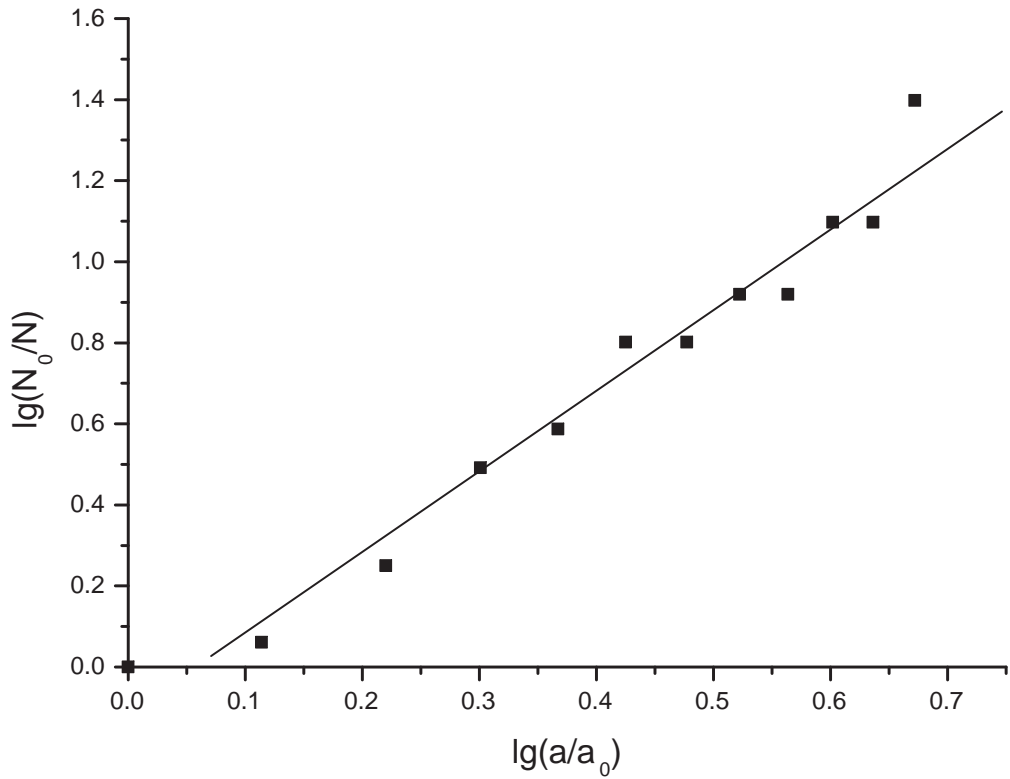}
\end{center}
\caption{The integral number of short pulses $N$ as a function of the amplitude $a$. North antenna data.
Linear fitting in logarithmic scale: $\log(N_0/N) = 1.99 
\log(a/a_0)-0.11$, $a_0 = 8$ mv, $N_0 = 2$.}
\label{amp}
\end{figure}

{\it Amplitude distribution} of the observed radio pulses is shown at
Fig.~\ref{hamp} and Fig.~\ref{amp}. In \cite{GDMZ} it was shown 
that the amplitude of radio signal $a$
is proportional to the number of secondary electrons $n_s$ and due to this
is proportional to the energy $\varepsilon_p$ of the cosmic ray particle
creating EAS \cite{Bel}.  One can see from Fig.~\ref{amp}
that according to our observational
data  the integral number $N$ of signals in high amplitude range
is falling down with amplitude as $a^{-\kappa}$, $\kappa\approx 2$.
It is in reasonable agreement with the well known cosmic ray integral
distribution ($N\propto\varepsilon^{-1.8}$). That coincidence  could be
considered
as {\it direct indication that the observed radio pulses are generated by
high energy particles crossing the thundercloud}.
Of course statistics of radio signals shown at Fig.~\ref{amp} is very poor.
We present the data to demonstrate only the new opportunities for
cosmic ray studies using thunderstorm radio emission. From the
integral distribution presented at Fig.~\ref{amp} and its comparison with our
numerical simulations one can deduce that the energies of cosmic ray
particles generating observed pulses lays in the interval $2
\times 10^{14} eV < \varepsilon_p < 2\times 10^{15} eV$.

\noindent {\it Electric current}

\begin{figure}
\begin{center}
\includegraphics[height=15cm]{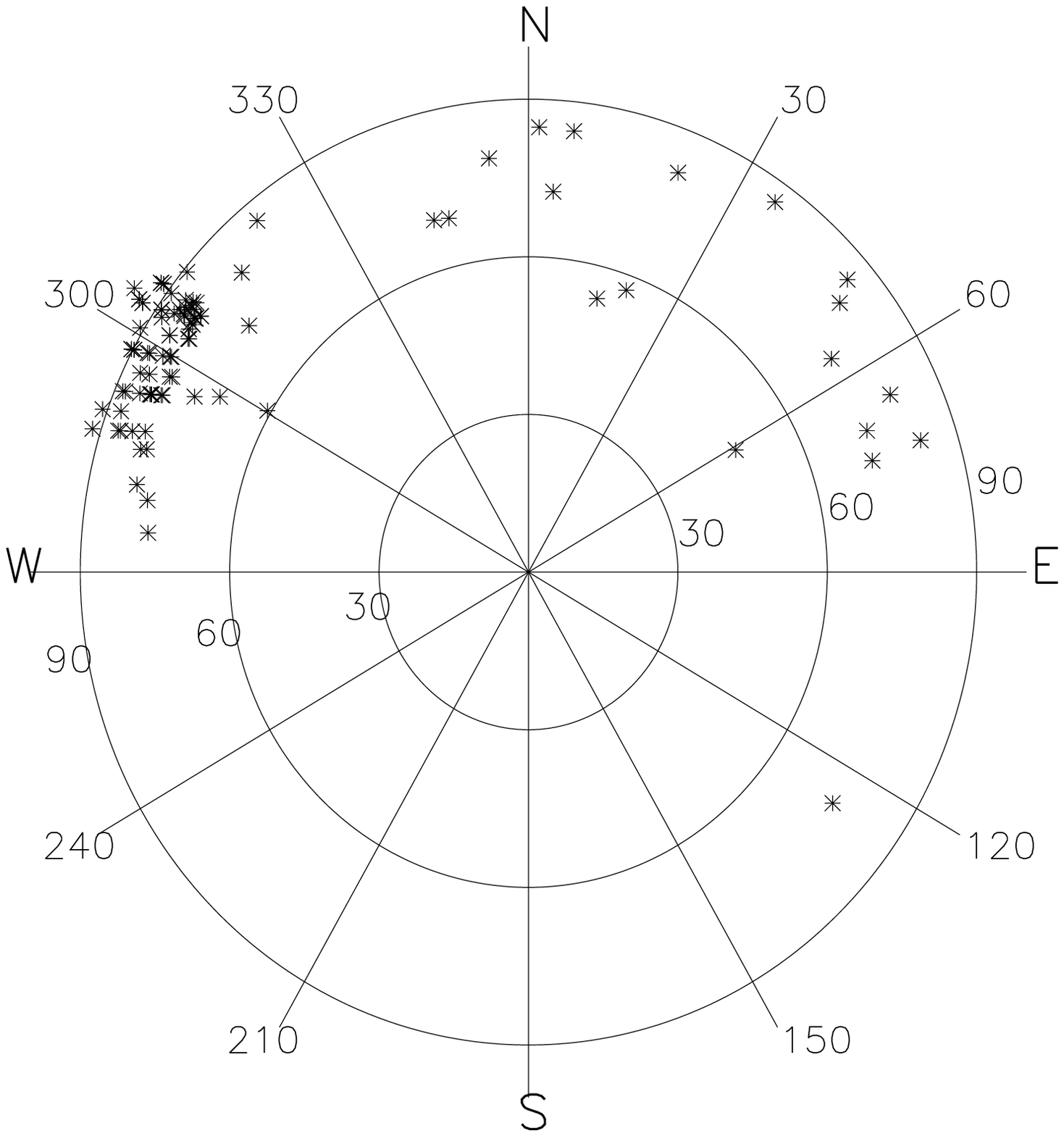}
\end{center}
\caption{Short radio pulses arrival angles during July 23, 2003 thunderstorm. 
Inclination angle is counted off from the zenith. 
Azimuth angle is counted off from the the North in the 
clockwise direction.}
\label{fig:rjul}
\end{figure}

\begin{figure}
\begin{center}
\includegraphics[height=15cm]{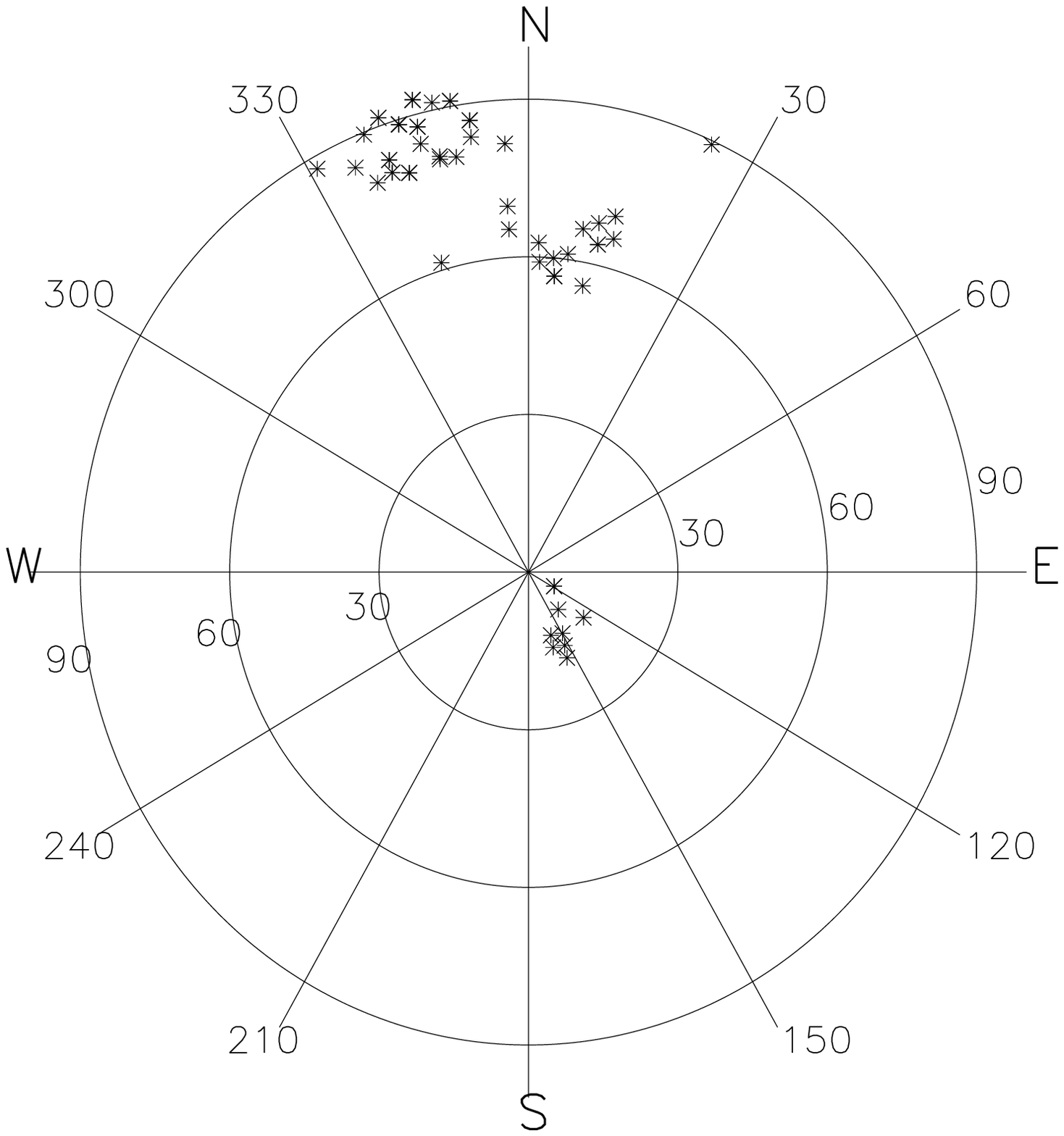}
\end{center}
\caption{Short radio pulses arrival angles during September 11, 
2003 thunderstorm. 
Inclination angle is counted off from the zenith. 
Azimuth angle is counted off from the the North in the 
clockwise direction.}
\label{fig:rsep}
\end{figure}

The main part of radio pulses had the bipolar form (Fig.~\ref{fig:record})
and is generated by the unipolar current $J$ directed vertically.
The pulses are coming mainly from horizontal direction (Figs.~\ref{fig:rjul},~\ref{fig:rsep}). For
the amplitude of radio pulse in this case it follows from \cite{GDMZ}
\begin{equation}\label{3a}
a = \frac{2 J}{c R}
\end{equation}
Here $R$ is the distance between current pulse generated in thundercloud
and trigger array. As it will be shown below $R\sim 1$ km.
From observations of pulses amplitudes presented in Fig.~\ref{hamp}
one can deduce that the maximal value $a_m$ corresponding to
cosmic ray particle energy $\varepsilon_p \approx 10^{15}$ eV is
\begin{equation}\label{4a}
a_m\approx 60 \hbox{mV}/\hbox{m}\,,\qquad\qquad
\varepsilon_p\approx 10^{15} \hbox{eV}.
\end{equation}
From (\ref{3a}), (\ref{4a}) it follows
\begin{equation}\label{J1}
J_a\sim \frac{\varepsilon_{15}}{R_{km}}\, A
\end{equation}
In this expression current is given in Amperes, $R_{km}$ -- the distance
from trigger array to RB region in kilometers and $\varepsilon_{15}$ --
the energy of CR particle which generates EAS (in $10^{15}$ eV).
Note that at the storm 23 July the radio pulses were in quite analogous
amplitude range  20 -- 40 mV/m but two pulses was extremely strong
$a\approx 200$ mV/m

\noindent {\it Full number of thermal electrons}

According to \cite{GMZ03} the current $J$ is produced by the motion of thermal
electrons ($\varepsilon \leq 1$ eV) under the action of RB electric field
$E\approx E_c\approx 2$ kV/cm:
\begin{equation}\label{J2}
J= e V_d n_T/l_p
\end{equation}
Here $e$ -- is the electron charge, $n_T$  -- full number of thermal electrons
at the characteristic length of a pulse $l_p$
and $V_d$  -- electron drift velocity:
\begin{equation}\label{J3}
V_d = \frac{e E}{m \nu}
\end{equation}
where $m$ -- electron mass, $\nu$ -- electron collision frequency in air
\cite{GBM},\cite{GNL}.
For $\varepsilon = 1$ eV and $N_m=1.7\times 10^{19} cm^{-3}$
(for thundercloud height 4 -- 5 km)

\begin{equation} \label{J4}
\nu = 10^{12} \left( \frac{N_m}{1.7\times 10^{19} cm^{-3}}\right) \, s^{-1}
\end{equation}

From (\ref{J1}) -- (\ref{J4}) it follows that

\begin{equation}\label{nT}
n_T = \frac{J l_p}{e V_d} \sim 10^{16} \frac{\varepsilon
_{15}}{R_{km}}
\end{equation}

\noindent {\it Dissipated energy}

Now let us take into account that for ionization of one air molecule
(generation  one free thermal electron) the average energy $\bar{\varepsilon
}= 20 \div 30$ eV is needed \cite{BAZ}. From (\ref{nT}) it follows that the
energy needed to generate the observed radio pulse is
\begin{equation}\label{E}
E\approx {\bar \varepsilon} n_T \sim 10^{17} eV
\end{equation}
Note that analogous energy goes to $\gamma$ - quanta generation.
Thus it follows from
(\ref{E}) that the {\it energy dissipated in RB - EAS interaction
region  at least about 100 times higher than the energy of CR
particle, generating EAS}! That is why we call the effect "Giant Burst".

Of course,  the dissipated energy comes from the interaction of
the high energy secondary RB - EAS electrons with thunderstorm electric field.
Gamma quanta, generating electrons and positrons can give support of the effect
also. The detailed  theoretical discussion of basic processes
which determine the Giant Burst is given in the
accompanying paper \cite{GDMZ2}.

\noindent {\bf II Thunderstorm Extensive Atmospheric Showers (TEAS)}

\noindent {\it Two main problems}

The trigger array (system I, II) worked stable during
thunderstorm with the same counting frequency as in quiet days:
one shower was registered by the joint system I-II every  2.45
sec (Fig.~\ref{evnum}). At the same time, the cases, when radio pulses were
observed, were singled out by two very significant
peculiarities:

1. {\it The inclination angles for most observed radio pulses were
large}:
$\theta \geq 60^{o} \div 90^{o}$ (Figs.~\ref{fig:rjul},~\ref{fig:rsep}, 
Tables 1,~2).

We note, that the previous detailed study of inclination angles
of EAS generated by the same energy CR particles ($\varepsilon \sim
10^{14} \div 10^{15}$) at Tien Shan Scientific Station gave quite a
contrary statements. For example according to \cite{BARS} the
probability of inclination angles deviation from vertical direction
$\theta = 0^o$ is proportional to $cos^6\theta$. It means that the
pulses should be concentrated near zenith.
Analogous result is obtained in \cite{AS}:
the full number $N$ of EAS with $\theta < 30^o$ was twice larger than
for the  $\theta > 30^o$:

\begin{equation}\label{fractN}
\frac{N(\theta < 30^o)}{N(\theta > 30^o)}\approx 2
\end{equation}
From (\ref{fractN}) it follows that the average angle density
distribution $\rho$ for
EAS with $\theta < 30^o$ is {\it fourteen times} higher than
for $\theta > 30^o$:

\begin{equation}\label{fractR}
\frac{\rho(\theta > 30^o)}{\rho(\theta < 30^o)} \approx
\frac{1}{2 \int_{30^o}^{90^o} sin\,\theta\,d \theta}
\int_{0}^{30^o}\sin\theta\,d\theta\approx 0.07
\end{equation}

Thus for large angles $\theta \sim 60^o \div 90^o$ the probability to
observe EAS under quiet conditions is very low, practically close to zero.

\begin{figure}
\begin{center}
\includegraphics[width=15cm]{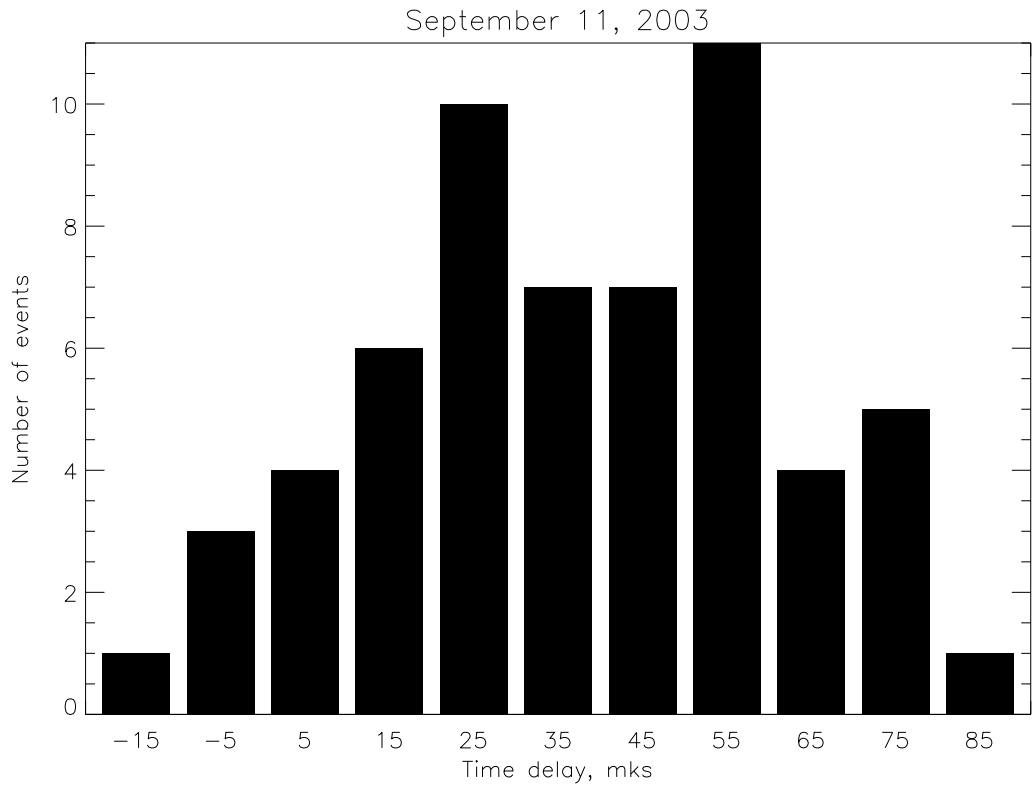}
\end{center}
\caption{Histograms of the delays between radio pulse arrival and EAS 
during September 11, 2003 thunderstorm. Time delay in $\mu$s.}
\label{fig:hsep}
\end{figure}

2. {\it The observed delay between a trigger array signal matching the EAS
coming and the radio pulse arrival to radio antenna was large}.
EAS was usually late and the time delay was quasi-random
distributed in the 5 -- 80 $\mu \mbox{s}$ timescale (Fig.~\ref{fig:hsep},Table~2).

We note that time delay between a EAS and radio signals should be no more
than trigger window $5 \mu \mbox{s}$, as both
signals are propagating with the light velocity.

Thus observed during thunderstorm specific effects are in direct contradiction
with the well established  usual observations of EAS in quiet conditions.

\noindent {\it Giant burst}

The qualitative explanation of these effects could be given
basing on the RB-EAS combined action analysis presented in the Part
I of this Section. Analyzing the
radio pulse production mechanism we have seen that when EAS pass
the thundercloud under RB conditions  fulfilled
a large amount of energy is dissipated due to the combined action of
thunderstorm electric field and cosmic ray secondaries.
This energy lead to the production of a giant
burst of electrons and $\gamma$-quanta. Electrons in their final thermal
state generate the observed radio pulse \cite{GMZ03},\cite{GDMZ}.
Gamma quanta determine
both peculiarities observed in Thunderstorm Extensive Atmosphere
Showers (TEAS).

\begin{figure}
\begin{center}
\includegraphics[width=10cm]{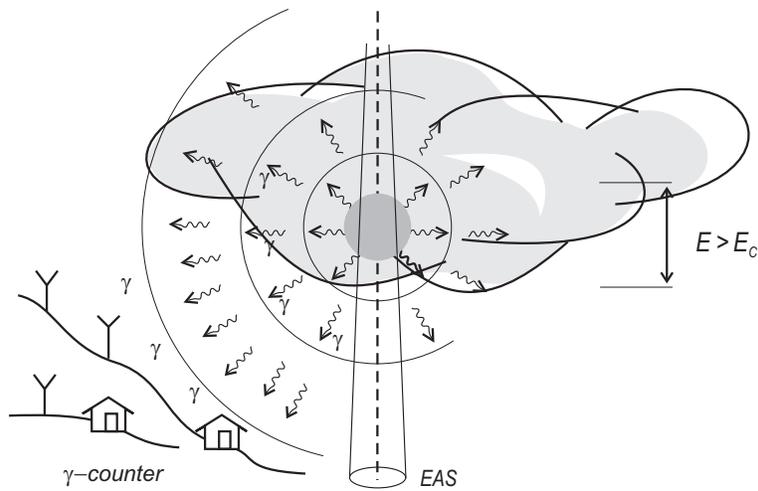}
\end{center}
\caption{Gamma burst.}
\label{loiko}
\end{figure}

Qualitatively it is shown at Fig.~\ref{loiko}. Giant number of high energy
electrons and $\gamma$-quanta generated in RB-EAS interaction region form a
cascade wich is moving in all directions. In the cascade due to
Compton inelastic scattering and $e^+e^-$ pair production on
electrons and air nucleas the $\gamma$-quanta energy is
rapidly diminishing. As soon as it
become less than 0.5 Mev $(\varepsilon_{\gamma} \leq mc^2)$
the motion of gamma quanta changes dramatically: the
energy losses are small now and the main process  become
quasy-elastic Compton scattering. As a consequence the motion of
$\gamma$-quanta with the energies $\varepsilon_{ph}
< \varepsilon_{\gamma} < mc^2$ is diffusive (below
$\varepsilon_{ph}=$20 keV photo-ionization losses dominates).
The diffusion is not a straightforward motion -- it lead
to the time delay $\Delta t$:

\begin{equation}\label{deltat}
\Delta t \approx R^2/2D_{\tau}, \quad
D_{\tau}=\frac{l_{\tau}c}{3}
\end{equation}

Here R is the distance $\gamma$-quanta passed by diffusion,
$l_{\tau}$ -- mean free path of $\gamma$-quanta determined by
Thomson scattering. In air $l_{\tau}=1/(\sigma_{\tau}ZN_m)
\approx 50 m$, where $\sigma_{\tau}$ -- Thomson crossection, $Z$ and
$N_m$ -- average number of electrons and number density in air
molecula. It follows from (\ref{fractR}) that for $R \approx 1
km$ the time dealy is $\Delta t\approx 100 \mu s$.

The proposed model explains both main peculiarities of TEAS:

\begin{enumerate}
\item  According the scheme at Fig.~\ref{loiko} the
trigger array  could see the pulses mainly at large inclination
angles. The direct EAS could  come to  trigger array rarely.
\item  It is natural to suppose in accordance with Fig.~\ref{loiko} that the
array was excited by strong gamma fluxes from giant bursts in
RB-EAS regions. It explaines the random time delay of shower
array signals up to $100 \mu$s (Figs.~\ref{fig:hjul},~\ref{fig:hsep}), 
as the $\gamma$-flux motion
was diffusive (\ref{deltat}).
\item Analysis shows that we registered not EAS 
itself but a gumma burts generated by RB-EAS process. That is 
why the energy of the primary cosmic ray particle could be 
larger than it was supposed previously.

More detailed theoretical analysis of the process 
is presented in the accompanying paper \cite{GDMZ2}.
\end {enumerate}

\noindent {\it Direct observations of EAS}

\begin{figure}
\begin{center}
\includegraphics[width=15cm]{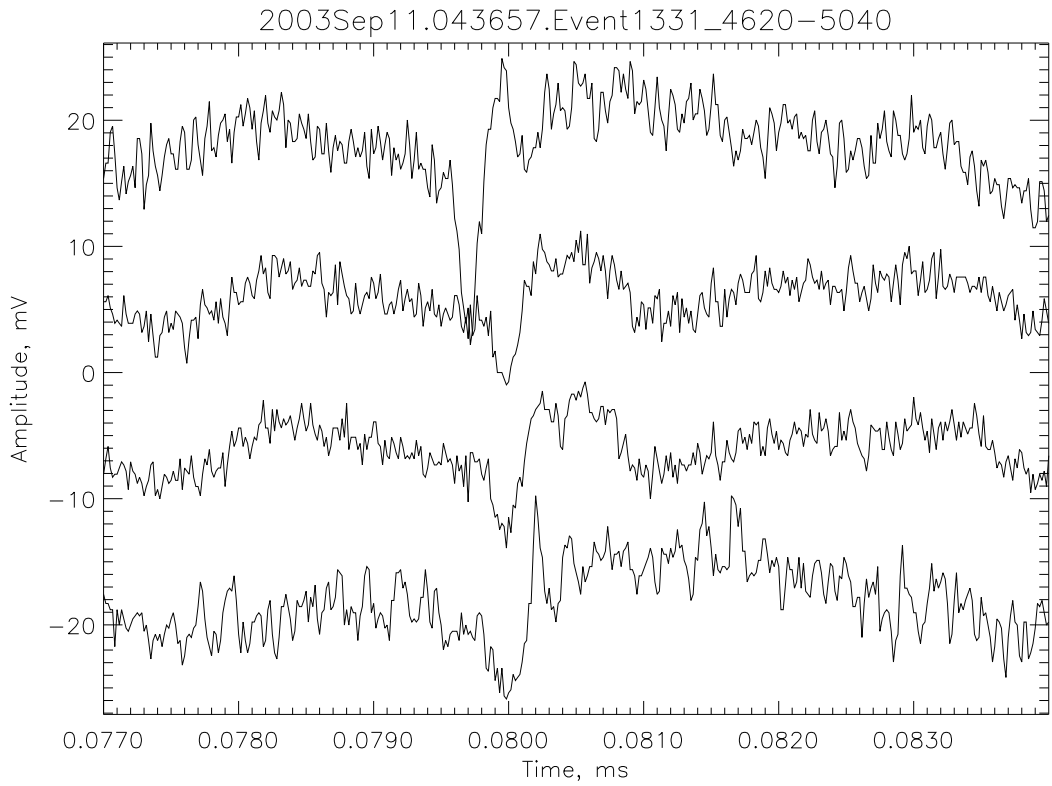}\\
\includegraphics[width=15cm]{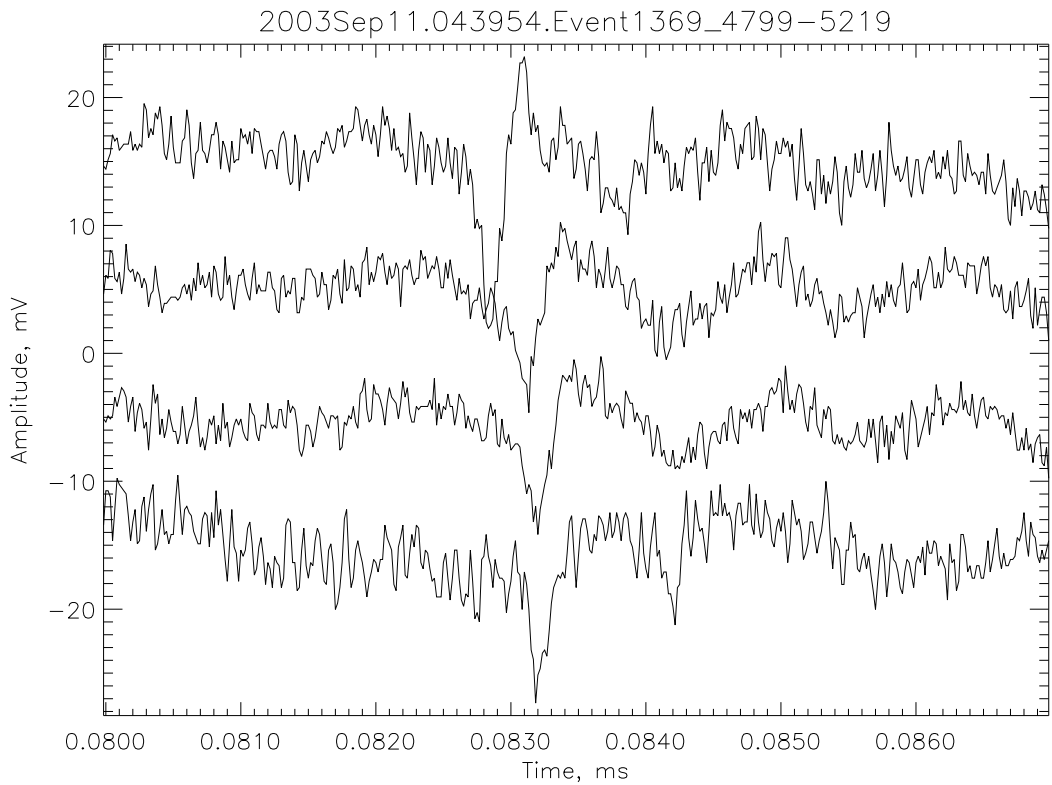}
\end{center}
\caption{Record of EAS pulses without time delay.}
\label{even1}
\end{figure}

In two events  the time delay between trigger and radio signal
is  small enough: ($0.2\,,3.3 \mu s$). It lays inside the trigger
array accuracy ($5 \mu s$).
These events shown in Fig.~\ref{even1} could be considered as the direct
simultaneous observation of radio pulses and EAS. We see that the
pulse records are practically the same as in usual case with
time delay.
The inclination angles for both events are
 large $\theta \approx 87^o,  89^o$,
what means practically the horizontal motion of CR particle.

\noindent {\it Near zenith direction}

\begin{figure}
\begin{center}
\includegraphics[width=15cm]{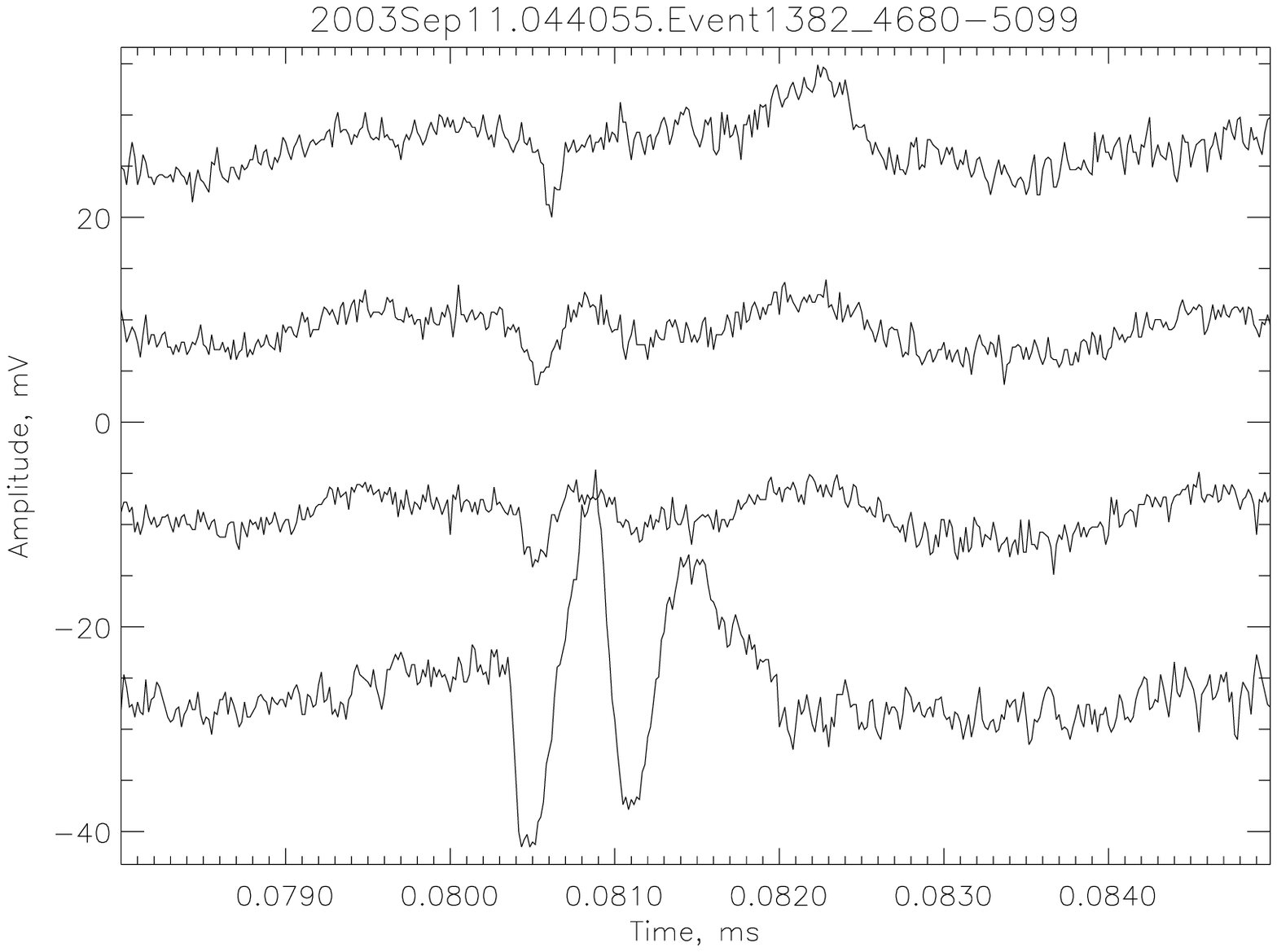}\\
\vspace{10mm}
\includegraphics[width=15cm]{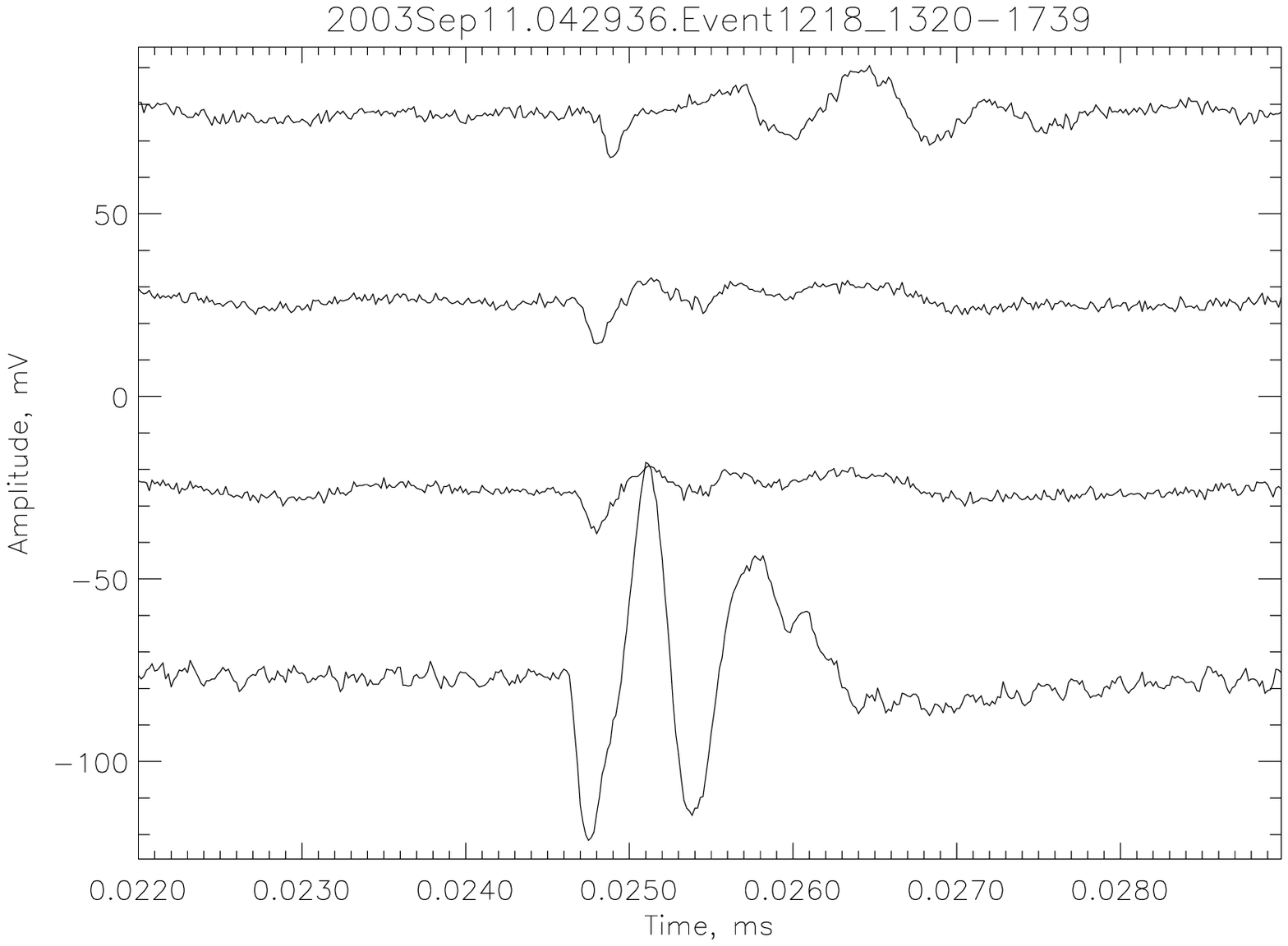}
\end{center}
\caption{Up -- a record of zenith EAS pulse without time delay, down -- 
a record of zenith EAS pulse with time delay}
\label{spec}
\end{figure}

From the data presented in Fig.~\ref{fig:rsep}, Table~2 one can see that nine
pulses are singled out -- they are directed close to zenith:
inclination angles from $5^o$ to $18^o$. They all are peculiar
pulses  being wider and having an unusual structure shown in 
Fig.~\ref{spec}. The radio pulses looks generated by double  unipolar current separated
in time at 0.5 $\mu$s or about 50 -- 100 m in space. Note that
one of the pulses presented in Fig.~\ref{spec} is a direct EAS pulse triggered
at time delay 2.7 $\mu$s which lies inside the 5$\mu$s trigger window.
Again no difference is seen between records of direct and delayed pulses.

We have no  clear understanding of the origin of these specific pulses.
The following preliminary explanation could be proposed. The current in
RB - EAS interference region has two components: the current of thermal
electrons and current of relativistic runaway electrons. The thermal
current is stronger and generates a wide spread radio emission. The
relativistic current is weaker and its radio emission is concentrated
near EAS direction \cite{33}. It is possible that in the close to zenith case
we observe the emission generated by both current pulses what can explain
its bipolar form. This speculation needs of course much
more detailed theoretical and experimental study.

\noindent {\it Lightning initiation}

In previous work \cite{G03} the lightning initiation process was studied using
analogous radio technique. The first lightning radio pulse was singled out
and observed for large lightning number. It was established that the
radio pulse is generated by unipolar current pulse and its form, width
and amplitude agree with the predictions of RB - EAS interaction theory
\cite{GDMZ}. Thus it was supposed that the lightning is initiated by combined
action of RB - EAS generated by primary cosmic ray particles with energy
$\varepsilon_p\geq 10^{16}$ eV. In the present work the radio pulses
and EAS are observed directly. The radio pulses have quite analogous form
and frequency range as in \cite{G03}. The difference in amplitudes is explained
by smaller energy of cosmic ray primary particle ($\varepsilon\approx 10^{15}
$ eV). Note that the observations of lightning first pulses
from the faraway thunderstorm at Tien Shan gave the results quite analogous
obtained in \cite{G03}.

Thus one can  state that the observations presented in the paper confirm
the lightning RB - EAS initiation mechanism proposed in \cite{GDMZ},
\cite{G03}.

Note that the radio pulses associated with extensive atmospheric showers
were observed in a number of previous works \cite{JEL} - \cite{ALEKS}.
The width and amplitude of observed pulses were quite different -- from
$5 \mu$s \cite{SUGA}
up to 100 $\mu$s \cite{ALEKS}, depending on conditions and technique used.
In \cite{KUSUK}, \cite{ALEKS} the connection
between the existence of radio  pulses and weather conditions
(clouds, thunderstorms) was indicated.

\section{Conclusions}

In conclusion we formulate briefly main results of the 
experiments described in the paper:

\begin{enumerate}
\item At Tien Shan Mountain Scientific Station a new EAS - Radio
installation is constructed. The installation consists of a wide 
spread EAS trigger array and a high time resolution radio
interferometer what allows to study the EAS 
and radio pulses emission simultaneously.

\item The absence of simultaneous EAS and radio pulses in a 
quiet (non-thunderstorm) atmosphere is established.

\item During two thunderstorms 150 simultaneous EAS and radio 
pulses were observed. The distribution of radio pulses on 
amplitude, form, inclination and azimuth direction was 
studied. The following main peculiarities are established:

\begin{enumerate}
\item The radio pulses came mainly from the horizontal direction.

\item There is a significant random time delay (up to 100 $\mu$s) 
between trigger signal and radio pulse.
\end{enumerate}

\item The analysis of the obtained experimental data allows to 
establish the existence of a new phenomena -- giant 
electron-gamma bursts generated by high-energy cosmic ray 
particles in thunderclouds. The burst energy could be $10^2 \div 
10^3$ times larger than the particle energy due to the RB-EAS 
interaction.

\item The burst generated by high energy cosmic ray particles 
$\varepsilon > 10^{16}$eV can serve for lightning initiation. That 
confirms the lightning initiation mechanism studied in \cite{G03}.
\end{enumerate}

The current generated in thunderclouds due to RB-EAS interaction 
is proportional to the cosmic ray particle energy 
$\varepsilon_p$. It means that the radio pulse amplitude $a$ for 
$\varepsilon_p \geq 10^{18} \div 10^{19}$ eV should be extremely high.
The pulse has definite characteristics quite analogous to the 
established in the present work. It could be observed from the 
immense distances -- 1000 km and even more. 

The further simultaneous observations of EAS together with giant 
bursts of X rays and radio pulses can serve for the development 
of a new effective method for the radio detection
of high energy cosmic ray particles.

\newpage
{\large {\bf Acknowledgements}}

The authors are grateful to Prof V.L.Ginzburg, Prof. 
E.L.Feinberg and Dr. H.Carlson foe useful discussions. The work 
was supported by EOARD-
ISTC grant \#2236, ISTC grant \#1480, by the President of Russian Federation 
Grant for Leading Scientific Schools Support
and by the Russian Academy Fundamental Research Program "Atmosphere Physics: 
Electric Processes, Radio Physics Methods".

\begin{table}
\label{tab:jul}
\caption{Short electromagnetic pulses recorded during July 23, 2003 thunderstorm.}
\begin{center}

{\tiny
\begin{tabular}{|c|c|c|c|c|}
\hline
  Time   & Event \# & Trigger delay & Inclination & Azimuth \\
\hline
17:51:32 &   2910   &       9.7     &      86     &    35   \\
17:57:33 &   2942   &      10.3     &      82     &   294   \\
17:57:50 &   2946   &       9.8     &      79     &   354   \\
17:59:14 &   2954   &      52.1     &      84     &   296   \\
         &          &      44.5     &      81     &   297   \\
18:02:06 &   2971   &      33.6     &      81     &   303   \\
18:02:20 &   2972   &      77.3     &      84     &   305   \\
18:02:40 &   2973   &      34.1     &      74     &   296   \\
         &          &      18.5     &      86     &   295   \\
18:17:34 &   3058   &      44.3     &      86     &   306   \\
         &          &      30.0     &      90     &   289   \\
18:19:02 &   3067   &      17.8     &      93     &   303   \\
18:19:50 &   3072   &      69.6     &      84     &   302   \\
18:24:04 &   3099   &      23.6     &      80     &   286   \\
18:27:18 &   3116   &      38.7     &      82     &   305   \\
18:29:36 &   3126   &      74.1     &      84     &   306   \\
18:30:08 &   3133   &      23.2     &      82     &   304   \\
18:30:33 &   3137   &      82.9     &      84     &    49   \\
         &          &      22.5     &      80     &    50   \\
18:35:49 &   3166   &      16.3     &      82     &   305   \\
         &          &       9.9     &      81     &   303   \\
18:37:49 &   3172   &      64.6     &      84     &   302   \\
         &          &      42.0     &      92     &   303   \\
18:40:13 &   3187   &       5.5     &      90     &   298   \\
18:40:36 &   3191   &      56.9     &      60     &   300   \\
18:40:55 &   3197   &      65.3     &      81     &   289   \\
18:47:49 &   3239   &     -11.8     &      89     &   304   \\
18:51:31 &   3267   &      45.9     &      86     &   320   \\
18:52:04 &   3270   &      71.4     &      80     &   294   \\
         &          &      64.3     &      70     &   298   \\
18:53:29 &   3279   &      73.3     &      87     &   298   \\
         &          &      49.3     &      86     &   288   \\
18:53:31 &   3280   &      61.3     &      83     &   293   \\
         &          &      47.7     &      92     &   303   \\
18:57:12 &   3300   &      38.6     &      47     &    60   \\
18:57:36 &   3303   &      62.6     &      84     &   307   \\
18:58:24 &   3312   &      32.4     &      82     &   294   \\
         &          &       5.3     &      89     &   298   \\
18:59:39 &   3323   &      80.6     &      82     &   294   \\
         &          &      48.5     &      80     &   281   \\
         &          &      26.3     &      76     &   275   \\
19:00:46 &   3328   &       1.3     &      86     &   304   \\
19:03:52 &   3340   &      48.2     &      86     &   288   \\
         &          &     -14.6     &      81     &   294   \\
19:05:30 &   3350   &      72.4     &      92     &   306   \\
19:06:07 &   3354   &       8.5     &      93     &   303   \\
         &          &      -0.1     &      81     &   306   \\
19:08:00 &   3364   &      35.4     &      86     &   298   \\
19:09:00 &   3374   &      27.2     &      82     &   305   \\
19:10:24 &   3380   &     -11.4     &      87     &   290   \\
19:10:25 &   3381   &       6.2     &      84     &   305   \\
19:11:46 &   3386   &      57.0     &      84     &   307   \\
19:12:13 &   3388   &      42.8     &      89     &   306   \\
19:12:55 &   3389   &      82.7     &      85     &   306   \\
         &          &      79.6     &      90     &   300   \\
19:13:02 &   3391   &      68.1     &      81     &   314   \\
19:15:23 &   3405   &      24.3     &      82     &   304   \\
19:16:10 &   3410   &      -6.6     &      88     &   292   \\
         &          &     -14.9     &      81     &   286   \\
19:24:28 &   3466   &      64.3     &      82     &   299   \\
19:25:05 &   3471   &     -11.5     &      92     &   306   \\
19:25:41 &   3476   &      67.5     &      89     &   298   \\
19:30:44 &   3512   &      50.6     &      91     &   306   \\
19:31:31 &   3521   &      35.0     &      53     &    14   \\
         &          &      -6.3     &      57     &    20   \\
19:32:58 &   3540   &      28.3     &      90     &   298   \\
19:34:38 &   3557   &      62.5     &      85     &   293   \\
         &          &     -12.6     &      80     &   297   \\
19:34:50 &   3559   &       8.4     &      77     &   280   \\
19:35:21 &   3568   &      34.5     &      81     &    21   \\
19:36:10 &   3578   &       4.4     &      88     &   303   \\
19:37:19 &   3582   &      24.8     &      95     &   304   \\
19:42:48 &   3623   &      61.7     &      72     &     3   \\
         &          &       2.8     &      84     &     1   \\
19:56:09 &   3756   &      -9.9     &      82     &    72   \\
         &          &     -12.8     &      73     &    68   \\
19:57:11 &   3766   &      11.6     &      75     &   125   \\
19:58:23 &   3775   &      35.7     &      80     &    65   \\
20:01:13 &   3800   &       5.0     &      73     &   309   \\
20:05:20 &   3828   &      13.0     &      82     &   299   \\
20:15:49 &   3897   &      25.7     &      73     &    56   \\
20:15:56 &   3898   &      47.6     &      83     &   288   \\
         &          &      24.9     &      89     &   304   \\
20:16:06 &   3899   &      41.9     &      72     &    72   \\
20:22:48 &   3946   &      51.0     &      81     &   306   \\
         &          &      13.3     &      89     &   304   \\
20:24:01 &   3953   &       8.3     &      87     &   293   \\
         &          &      -0.9     &      91     &   306   \\
20:30:37 &   4002   &      26.3     &      83     &   306   \\
         &          &      16.6     &      89     &   309   \\
20:32:37 &   4010   &      47.9     &      69     &   344   \\
         &          &      45.4     &      69     &   346   \\
20:33:13 &   4016   &       5.9     &      84     &     6   \\
20:33:39 &   4019   &      23.4     &      84     &   299   \\
20:37:10 &   4052   &      55.0     &      86     &   305   \\
         &          &      31.8     &      82     &   305   \\
20:41:34 &   4074   &      44.0     &      82     &   294   \\
20:43:14 &   4087   &      19.2     &      91     &   287   \\
\hline
\end{tabular}}

\end{center}
\end{table}

\begin{table}
\label{tab:sep}
\caption{Short electromagnetic pulses recorded during September 11, 2003 thunderstorm.}
\begin{center}

{\tiny
\begin{tabular}{|c|c|c|c|c|}
\hline
  Time   & Event \# & Trigger delay & Inclination & Azimuth \\
\hline
04:26:32 &   1180    &     75.0     &      69     &    14   \\
04:28:11 &   1199    &     72.0     &       5     &   117   \\
         &           &     50.6     &       9     &   140   \\
04:28:58 &   1210    &     54.2     &      81     &   347   \\
04:29:03 &   1212    &     57.6     &      81     &   356   \\
04:29:13 &   1214    &     72.7     &      15     &   160   \\
04:29:36 &   1218    &     58.4     &      14     &   128   \\
04:30:08 &   1225    &     50.8     &      63     &    12   \\
04:30:40 &   1233    &     30.1     &      67     &    12   \\
04:32:12 &   1258    &     32.2     &      91     &   347   \\
04:32:20 &   1261    &     40.2     &      62     &     1   \\
04:33:23 &   1275    &     47.6     &      65     &   356   \\
04:36:50 &   1327    &     55.0     &      12     &   159   \\
         &           &    -11.6     &      18     &   154   \\
04:36:57 &   1331    &      3.3     &      86     &   352   \\
04:36:59 &   1332    &     60.1     &      55     &    11   \\
04:38:44 &   1354    &     29.3     &      83     &   352   \\
04:39:54 &   1369    &      0.2     &      87     &   345   \\
04:39:58 &   1370    &     14.9     &      56     &     5   \\
04:40:39 &   1379    &     49.6     &      89     &   343   \\
04:40:53 &   1381    &     20.3     &       5     &   117   \\
04:40:55 &   1382    &      2.7     &      13     &   149   \\
04:42:17 &   1405    &     30.3     &      66     &     9   \\
04:42:34 &   1410    &     27.8     &      61     &     7   \\
04:43:19 &   1425    &     38.2     &      86     &   352   \\
04:43:30 &   1428    &     70.8     &      83     &   340   \\
04:44:10 &   1440    &     35.7     &      69     &   356   \\
04:14:26 &   1442    &     49.4     &      83     &   340   \\
04:44:55 &   1447    &     53.1     &      87     &   345   \\
04:45:06 &   1450    &     22.6     &      80     &   349   \\
04:45:37 &   1456    &     60.3     &      92     &   345   \\
04:45:50 &   1458    &     19.3     &      15     &   152   \\
04:46:02 &   1462    &     -9.6     &      89     &   343   \\
04:46:07 &   1463    &     20.5     &      92     &   345   \\
         &           &      4.5     &      59     &     4   \\
04:46:51 &   1475    &     17.2     &      91     &   350   \\
04:46:54 &   1477    &     48.6     &      89     &    24   \\
04:47:07 &   1480    &     28.3     &      65     &    15   \\
04:47:16 &   1483    &     36.6     &      80     &   337   \\
04:47:56 &   1489    &     -5.9     &      84     &   345   \\
04:47:57 &   1490    &     -1.2     &      89     &   343   \\
04:47:58 &   1491    &     16.3     &      80     &   340   \\
04:48:31 &   1501    &     54.5     &      61     &   343   \\
04:48:54 &   1508    &     26.7     &      87     &   331   \\
04:48:56 &   1509    &     66.0     &      84     &   335   \\
04:49:13 &   1514    &     70.5     &      89     &   343   \\
04:49:15 &   1515    &     18.4     &      92     &   345   \\
04:49:16 &   1516    &     82.4     &      80     &   347   \\
04:49:40 &   1520    &     24.6     &      59     &     2   \\
04:51:13 &   1542    &     59.8     &      91     &   340   \\
04:51:51 &   1557    &     45.8     &      91     &   340   \\
04:52:06 &   1560    &     26.3     &      79     &   342   \\
04:53:31 &   1583    &     62.6     &      80     &   340   \\
04:53:59 &   1598    &     40.6     &      56     &     5   \\
04:54:26 &   1606    &     35.6     &      63     &    12   \\
04:57:47 &   1657    &     54.0     &      92     &   345   \\
04:58:05 &   1663    &     52.3     &      91     &   350   \\
05:08:18 &   1849    &     17.9     &      79     &   342   \\
05:21:47 &   2079    &     28.0     &      89     &   338   \\
\hline
\end{tabular}}

\end{center}
\end{table}

\newpage


\begin{thebibliography}{50}
 

\bibitem{GMR} A.V.Gurevich, G.M.Milikh, R.Roussel-Dupre Phys. 
Lett. A {\bf 165}, (1992) 463.

\bibitem{Wil} C.T.R.Wilson  Proc. Cambridge Philos.Soc. {bf 22}, 
(1924) 34.

\bibitem{RD} R.Roussel-Dupre, et al Phys.Rev. E {\bf 49} (1994) 2257.

\bibitem{L} N.G.Lehtinen, et al. Geophys. Res. Lett. {\bf 24} 
(1997) 2639.

\bibitem{B} L.P.Babich, et al. Phys.Lett.A {\bf 245}, (1998) 460. 

\bibitem{G1} A.V.Gurevich, et al.  Phys. Lett. A {\bf 275} 
(2001) 101.

\bibitem{UFN} A.V.Gurevich, K.P.Zybin  Physics Uspekhi {\bf 44}, 
11 (2001) 1119.

\bibitem{GMZ03} A.V.Gurevich, Yu.V.Medvedev, K.P.Zybin  Phys. 
Lett. A {\bf 321} (2004) 179. 

\bibitem{Marsh} T.Marshall et al. J. Geophys. Res. {\bf 100} 
(1996) 7097.

\bibitem{MacG} D. MacGorman, W.D.Rust The electrical nature of the storms,
New York, Oxford Univ. Press (1998)

\bibitem{AVG} A.V.Gurevich, et al.  Phys. Lett. A {\bf 282}, 180, (2001)

\bibitem{DWY} J.R.Dwyer Geophys. Res. Lett. {\bf 30} (2003), 2055.

\bibitem{McC} Mc Carthy, G.Parks Geophys.Res.Lett {\bf 12} (1985) 393. 

\bibitem{Eack} K.B.Eack, et al. J.Gophys.Res. {\bf 101} (1996) 29637.

\bibitem{Ch} A.P.Chubenko, et al. Phys.Lett.A. {\bf 275} (2000) 90.

\bibitem{BRU} M.B.Brunetti, et al. Geophys. Res. Lett. A {\bf 309} (2000)
1599.

\bibitem{ALEX} V.V. Alexeenko, et al. Phys. Lett. A {\bf 301} (2002) 299.


\bibitem{Ch2} A.P.Chubenko, et al. Phys.Lett.A. {\bf 309}(2003) 90.

\bibitem{18a} T.Torii et al JGR {\bf 107} (2002) 4324

\bibitem{GDMZ2} A.V.Gurevich, Yu.V. Medvedev and K.P.Zybin 
(2004) (to be published in arXiv:hep-ph).

\bibitem{GDMZ} A.V.Gurevich, et al. Phys. Lett A {\bf 301}, 
(2002) 320.

\bibitem{G03} A.V.Gurevich, et al. Phys. Lett. A {\bf 312} (2003) 228.

\bibitem{BARS} K.V.Barshakov, et al. Preprint FIAN {\bf \#19} (1998).

\bibitem{AS} V.S.Aseikin, et al. Proc FIAN {\bf v.19} (1979) 3.

\bibitem{GBM} A.V.Gurevich, N.D.Borisov, G.M.Milikh Physics of Microwave Discharges,
 Gordon and Breach, Amsterdam, 1997.

\bibitem{BAZ} E.M.Bazelyan, Yu.P.Raizer, Lightning Physics and 
Lightning Protection, IOP, Bristol, 2000.

\bibitem{Bel} S.Z.Belenkij Avalanche Processesin Cosmic Rays, OGIZ, 
Moscow (1948).

\bibitem{GNL} A.V.Gurevich Nonlinear Phenomena in the Ionosphere, Springer, New-York, 1978.

\bibitem{JEL} J.V.Jelly, et al. Nature, {\bf 205} (1965) 327.

\bibitem{SUGA} K.Suga, et al. 19 ICRC {\bf 7} (1985) 268.

\bibitem{KUSUK} M.Kusukuse, et al. 22 ICRC {\bf 4} (1991) 359.

\bibitem{ALEKS} A.V.Aleksandrov, et al. 20 ICRC {\bf 6} (1987) 132.

\bibitem{33} R.Roussel - Dupre, A.V.Gurevich JGR {\bf 101} (1996)
2297

\end{thebibliography}
\end{document}